\documentclass[9pt,twocolumn,twoside,dvipsnames]{pnas-new}

\templatetype{pnasresearcharticle} 

\newbox{\orcidlogo}
\sbox{\orcidlogo}{\large\includegraphics[height=1.8ex]{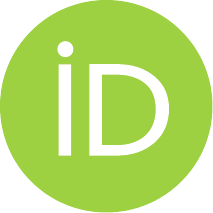}}
\newcommand{\orcid}[1]{\href{https://orcid.org/#1}{\usebox{\orcidlogo}\,}}

\title{Fast Iteration of Spaced \textit{k}-mers}

\author[]{Lucas Czech \orcid{0000-0002-1340-9644}}

\affil[]{Section for GeoGenetics, Globe Institute, University of Copenhagen, Denmark; \href{mailto:lucas.czech@sund.ku.dk}{lucas.czech@sund.ku.dk}}

\newcommand{\mainauthor}{}
\leadauthor{\mainauthor}

\setboolean{displaywatermark}{false}

\usepackage{listings}
\usepackage{lastpage}
\usepackage{verbatimbox}

\usepackage{caption}
\usepackage[labelformat=simple]{subcaption}

\usepackage[binary-units=true]{siunitx}

\renewcommand{\thesubsection}{\arabic{section}.\arabic{subsection}}

\newcommand{\figref}[1]{Fig.~\ref{#1}}

\usepackage{tikz}

\begin{abstract}
\mdseries
\textbf{Background:}
Short sequence substrings of a fixed length \textit{k}, called \textit{k}-mers, are a ubiquitous computational primitive in bioinformatics, used across sequence indexing, read mapping, genome assembly, metagenomic classification, and comparative genomics.
Spaced \textit{k}-mers generalize this concept by selecting only a subset of positions within a \textit{k}-mer, improving robustness to mismatches and sequencing errors.
While \textit{k}-mers are computationally highly efficient, spaced \textit{k}-mers require additional work to be extracted from a sequence, which has slowed down existing methods.
\\
\textbf{Results:} 
We present a collection of efficient algorithms for extracting spaced \textit{k}-mers from nucleotide sequences, optimized for different hardware architectures. They are based on bit manipulation instructions at CPU level, making them both simpler to implement and up to an order of magnitude faster than existing methods. We further evaluate common pitfalls in \textit{k}-mer processing, which can cause substantial inefficiencies. 
\\
\textbf{Conclusions:} 
Our approaches allow the utilization of spaced \textit{k}-mers in high-performance bioinformatics applications without major performance degradation compared to regular \textit{k}-mers, achieving a throughput of up to 750MB of sequence data per second per core.
\\
\textbf{Availability:} The implementation in C++20 is published under the MIT license, and freely available at
\href{https://github.com/lczech/fisk}{github.com/lczech/fisk}
\vspace{-3.5em}
\end{abstract}

\begin{document}

\verticaladjustment{-2pt}

\maketitle
\thispagestyle{fancy}

\ifthenelse{\boolean{shortarticle}}{\ifthenelse{\boolean{singlecolumn}}{\abscontentformatted}{\abscontent}}{}



\begin{figure*}[b]
    \centering
    \includegraphics[scale=0.63]{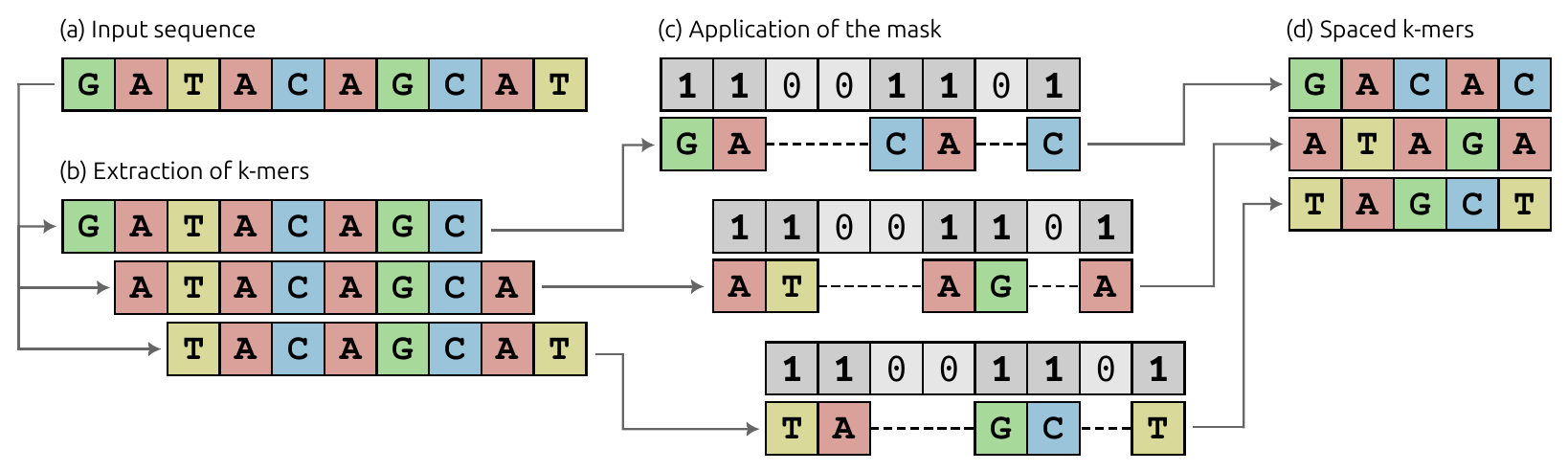}
    \begin{subfigure}{0pt}
        \phantomcaption
        \label{fig:Overview:sub:Input}
    \end{subfigure}
    \begin{subfigure}{0pt}
        \phantomcaption
        \label{fig:Overview:sub:Kmers}
    \end{subfigure}
    \begin{subfigure}{0pt}
        \phantomcaption
        \label{fig:Overview:sub:Masking}
    \end{subfigure}
    \begin{subfigure}{0pt}
        \phantomcaption
        \label{fig:Overview:sub:SpacedKmers}
    \end{subfigure}
    \vspace*{-1.5em}
    \caption{
        \textbf{Spaced \textit{k}-mer extraction.}
        \textbf{(a)} An input sequence of nucleotide characters.
        \textbf{(b)} Three consecutive $k$-mers of length $8$ are extracted from the sequence, each overlapping by $k-1 = 7$ characters, forming a sliding window of length $k$ over the sequence.
        \textbf{(c)} A fixed mask $m$ of span $k=8$ and weight $w=5$ is applied to each $k$-mer, to select characters where the mask is $1$, while leaving out characters where the mask is $0$.
        \textbf{(d)} The final spaced $k$-mers, which we here call $w$-mers for clarity.
    }
\label{fig:Overview}
\end{figure*}

\section{Background}
\label{sec:Background}

A $k$-mer is a contiguous substring of length $k$ extracted from a biological DNA or amino acid sequence. 
Due to their simplicity and efficiency, $k$-mers are a fundamental building block in computational genomics and bioinformatics \cite{Marchet2024-rd,Marchet2024-ro}, with applications ranging from genome assembly and read mapping to metagenomic classification and alignment-free sequence comparison \cite{Moeckel2024-fq}. For the DNA nucleotide alphabet $\left\{ \text{A}, \text{C}, \text{G}, \text{T} \right\}$, an example is shown in \figref{fig:Overview}(a-b).

Although contiguous $k$-mers are computationally efficient to extract, they are sensitive to mismatches. Many methods expect exact matches of $k$-mers, for instance, to index and search genomic sequences. Thus, a single nucleotide substitution disrupts all $k$-mers spanning the mismatching position. To improve robustness to mutations and sequencing errors, spaced $k$-mers were introduced \cite{Burkhardt2001-bb,Ma2002-hv}, where a fixed mask (or pattern) of length $k$ is applied to select a subset of positions in a $k$-mer, as shown in \figref{fig:Overview}(c-d). The resulting subsequence captures a context of span $k$, while allowing to tolerate mismatches at the masked (not selected) positions. Carefully designed and optimized patterns can substantially improve sensitivity in homology search and sequence comparison compared to contiguous $k$-mers \cite{Ma2002-hv,Hahn2016-gw,Kucherov2006-pm,Noe2017-cq}.

Spaced $k$-mers have since been adopted in a broad range of bioinformatics applications. They achieve a favorable balance between sensitivity and specificity, but also introduce computational challenges; previous work has hence focused on algorithms for efficient spaced $k$-mer extraction \cite{Girotto2018-wt,Girotto2018-ns,Petrucci2020-zt,Mian2024-nf,Gemin2026-du}.

Here, we present methods to extract spaced $k$-mers over the DNA nucleotide alphabet and with $k \leq 32$, which is the most common use case for tasks such as metagenomic classification and short read mapping; adaptation to other alphabets and extension to larger values of $k$ is possible. Compared to existing methods, our approaches are both faster and simpler to implement.

\section{Methods}
\label{sec:Methods}

We here give an overview of our algorithms; for details, see the Supplementary Text. 
The C++ code for this project is freely available under the MIT License at \href{https://github.com/lczech/fisk}{github.com/lczech/fisk}.
Our implementation is concise and well documented, in order to be easily adaptable; for a more complete suite of $k$-mer-related and other high-performance genomics functionality, see our C++ library genesis \cite{Czech2020-xt}.


\subsection{Problem Statement}
\label{sec:Background:sub:ProblemStatement}

DNA sequences consist of four nucleotides from the alphabet $\mathrm{N} = \left\{ \text{A}, \text{C}, \text{G}, \text{T} \right\}$. A $k$-mer $s$ is a sequence of length $k$ of characters from this alphabet, i.\,e., $s \in \mathrm{N}^k$. This is often represented via a compact 2-bit encoding, where $\text{A} = 00, \text{C} = 01, \text{G} = 10, \text{T} = 11$. This allows up to 32-mers to be stored in a single 64-bit machine word, thus enabling efficient storage and fast processing of $k$-mers via simple bit-wise operations on modern computer hardware. 
The application to other encodings is straightforward.

A spaced $k$-mer is defined via a binary mask $m \in \{0,1\}^k$ of length $k$ to select certain positions of an underlying $k$-mer (1s), while skipping others (0s), see \figref{fig:Overview}(c-d). We call $k$ the span of the mask, and $w$ its weight, with $w = \sum_i m_i$ the number of positions in the mask that are set to $1$. The resulting spaced $k$-mer is here called a $w$-mer (as it has length $w$), to distinguish it from the underlying $k$-mer it was extracted from.

The problem can then be stated as follows. Given a DNA nucleotide sequence of length $n$ over the alphabet $\mathrm{N}$, slide a window of size $k$ across the sequence, yielding $n - k + 1$ windows; in each window, apply the mask, by extracting the corresponding bases where $m$ is set to $1$ to construct a $w$-mer. 
As in \figref{fig:Overview}(d), the resulting $w$-mers thus only share short sub-strings with each other, namely where blocks of consecutive 1s overlap in the mask $m$ when shifted. This makes it difficult to ``re-use'' information across $w$-mers, which motivates the development of efficient techniques for this task.

Remarks on terminology: In the existing literature, the mask $m$ and the extracted $w$-mer are both called ``seed'' or ``spaced seed''. This is however overloaded and ambiguous, as the term ``seed'' usually refers to a value used to initialize an algorithm. 
Thus, the name ``(spaced) seed'' makes sense for (spaced) $k$-mers that are used in a seed-and-extend approach, but not in the general case, and not for the mask itself. Further names for the mask $m$ include ``pattern'', ``shape'', and ``template''. Existing methods also often use (spaced) $k$-mers as keys for lookups in hash tables; hence, the term ``$k$-mer hash'' has been used, where the encoded $k$-mer is a perfect hash value of itself. However, a ``hash'' is usually the result of a hash function, which takes arbitrary-length input instead of a fixed size $k$. We thus refer to the 2-bit representation as the ``encoding'' of the $k$-mer instead. 


\subsection{Encoding and \textit{k}-mer Extraction}
\label{sec:Methods:sub:EncodingExtraction}

The initial steps are to traverse the input sequence, transform the ASCII-encoded nucleotides into 2-bit encoding, and construct and iterate over the $k$-mers. There are several pitfalls in these steps that lead to major performance bottlenecks, which occasionally appear in published code. First, the encoding can be implemented as a switch-statement or series of if-statements, individually checking for the four nucleotides in $\mathrm{N}$. The latter in particular, however, generally invokes several CPU branch prediction misses per input character. Second, each $k$-mer can be fully re-extracted based on the underlying sequence, by applying the encoding $k$ times per $k$-mer, which is suboptimal.

The more optimized approach is to use a branchless encoding function, and to apply bit shifts between consecutive $k$-mers, thus re-using all overlapping bits from the previous $k$-mer in the sequence, and only encoding one new character in each iteration, as shown in \figref{fig:Overview}(b).

We tested two variants of branchless character encoding. The first is based on bit operations on the input character, and exploits a serendipity in the ASCII code to obtain the 2-bit encoding; the technique was recently independently discovered and utilized for bit-compressing genomic sequences \cite{Corontzos2024-kj}. The second is a simple lookup table containing the 2-bit codes for each single-byte character; this was more performant in our benchmarks, and is thus used henceforth.

Unless the input has previously been sanitized, there is one unavoidable conditional branch in the hot loop, to check if the current $k$-mer exclusively comprises valid characters. However, in typical data with but a few invalid characters that are not in $\mathrm{N}$, the CPU branch prediction mostly eliminates the overhead for this.

This yields a performance of $\approx$1\,ns per extracted $k$-mer, or $\approx$1\,GB of sequence data per second. We also tested a SIMD-accelerated implementation based on encoding 32 nucleotides at once with the ASCII exploit, using the AVX2 instruction set, which yielded a performance of $\approx$0.8\,ns per $k$-mer. Due to the data dependency between consecutive $k$-mers, this is difficult to further optimize; see the Supplementary Text for details.

Note that we limited our implementation to $k \leq 32$, for both the initial and the extracted spaced $k$-mer. For common applications such as homology search, metagenomic classification, and short read mapping this is the typically sufficient limit anyway, see Supplementary Table S1. This allows us to leverage rolling encoding, as a $k$-mer fits into a 64-bit word. Larger values of $k$ are however possible with additional bit operations or adaptation to, e.\,g., 128-bit registers, but left as future work. 





\subsection{Spaced \textit{k}-mer Extraction}
\label{sec:Methods:sub:SpacedKmerExtraction}

The main step then is the extraction of the spaced $k$-mers, i.\,e., the construction of the $w$-mers, at each position along the sequence. The ``naive'' extraction of a $w$-mer simply loops over the set of positions in the mask, and encodes each corresponding character from the input anew. Obviously, this is inefficient, as it requires a loop with $w$ iterations per $w$-mer.

The more efficient approach is to leverage the shift-based traversal of $k$-mers, and extract the $w$-mers directly from there. That is, given a $k$-mer in 2-bit encoding, and stored in a 64-bit word, we need to extract the pairs of bits corresponding to the positions that are set in the mask, and densely pack them towards the end of the word. Each bit pair then needs to be shifted according to the number of gaps in the mask between its position in the $k$-mer and the end of the mask, thus closing the gaps, as in \figref{fig:Overview}(c-d).

This operation is called \texttt{bit extract}. It gathers all bits denoted in the mask towards one end of the word and zeros out the rest. This is also known as bit compression, bit gathering, or bit packing, and is a well-understood operation with several dedicated implementations in software and hardware, which we are exploiting here. To our knowledge, it has not yet been applied to extract spaced $k$-mers.

On Intel CPUs (since 2013), as well as recent AMD CPUs (since 2020), the \texttt{PEXT} intrinsic implements bit extraction in hardware with high throughput; in our benchmarks, it is the most performant way to extract $w$-mers with but a few cycles of overhead compared to regular $k$-mer iteration. Older AMD CPUs (since 2015) also offer this intrinsic, but implement it via slow microcode. On ARM architecture (Apple CPUs), it is not available.

To mitigate this, and offer a fallback solution, we tested several algorithms to perform bit extraction in software; see the Supplementary Text for details. In our benchmarks, with no substantial exceptions, the most efficient implementation is an existing  algorithm \cite{Warren2012-sy} that requires $\log_2(b)$ steps for $b$ bits, e.\,g., 6 steps for 64-bit words. In each step, selected bits of the input are shifted in increasing powers of two, such that any gap length in the mask can be expressed as a combination of such shifts. For instance, a bit with 42 gaps (unset bits) towards the end of the mask is shifted in steps 2, 4, and 6, corresponding to shifts of 2, 8, and 32 bits, respectively. The shifts required for each bit are solely dependent on the mask, and can be pre-computed in a small table. Here, we call this the ``Butterfly'' algorithm, inspired by bit manipulation techniques that move bits across a machine word, where the pattern of movement resembles the wings of a butterfly. As the steps of the Butterfly algorithm are fixed for a given mask, they can be applied to several $k$-mers simultaneously. We thus implemented SIMD-accelerated variants of this for the SSE2, AVX2, AVX512, and ARM Neon intrinsics, each computing one consecutive $w$-mer per 64-bit lane.

Moreover, in many applications that rely on spaced $k$-mers, multiple masks are applied, e.\,g., to increase the sensitivity of sequence searches \cite{Ounit2016-hi,Hahn2016-gw}. In that case, the encoding and $k$-mer iteration only need to be executed once, and the $w$-mers for each mask can be extracted from that $k$-mer. We also implemented this with SIMD acceleration.


On older AMD CPUs, the \texttt{PEXT} intrinsic is available, but rather inefficient. This unfortunately prohibits us from simply testing for its presence at compile time to test if it can be used.
We hence implemented dynamic selection mechanisms at run time, where for a given mask a short benchmark is run to select the fastest available algorithm. Using C++ templates (or Rust generics, if ported), this mechanism can be used to switch between algorithms without overhead for function indirection.


\section{Results}
\label{sec:Results}

We here summarize the performance benchmarks; for details, see the Supplementary Text. 
All benchmarks are based on a sequence of randomly generated nucleotide data, using the same set of masks as the existing method MISSH \cite{Mian2024-nf}. The performance of the algorithms does generally not depend on sequence content; random data is thus valid for benchmarking here. 


\subsection{Benchmarks}
\label{sec:Results:sub:Benchmarks}

Results for several CPU architectures are shown in \figref{fig:ExtractSpacedKmers}; extended results are presented in the Supplementary Text. In short, both the Intel Xeon and the AMD Epyc offer fast \texttt{PEXT}, while the AMD Ryzen and Apple M1 do not. For every hardware architecture tested, at least one of our implementations has a throughput of one spaced $k$-mer per 1.3--1.8\,ns per core, equivalent to 520MB to 750MB per second for traversing a sequence and extracting its $w$-mers. This is more than an order of magnitude faster than the naive approach for extracting spaced $k$-mers. Compared to the baseline of $\approx$1\,ns for regular (contiguous) $k$-mer extraction from a sequence, this is likely fast enough for most applications, which usually incur a more significant overhead on top for their downstream computations.

\begin{figure*}[!htb]
    \centering
    \includegraphics[width=0.49\linewidth]{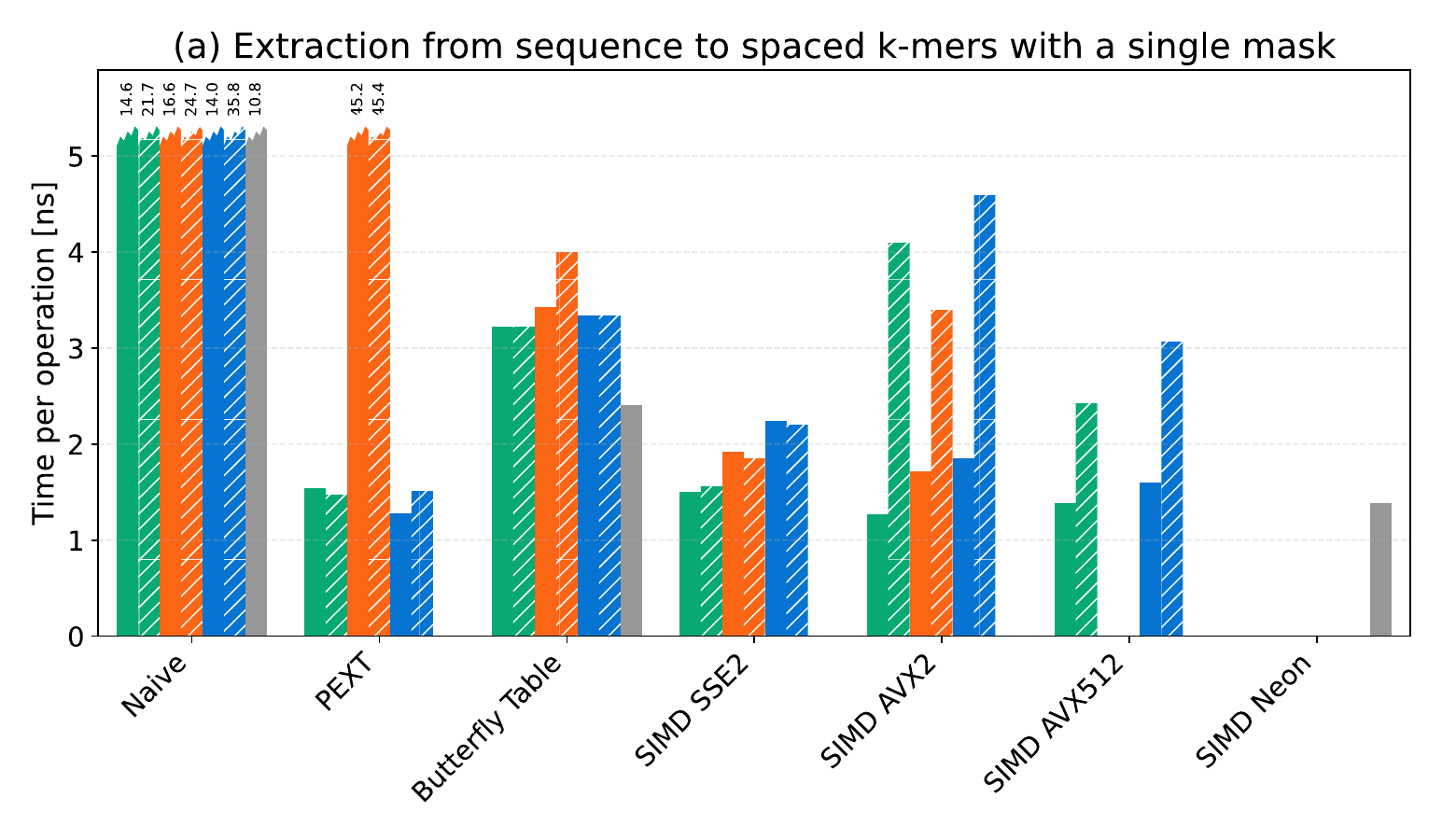}
    \includegraphics[width=0.49\linewidth]{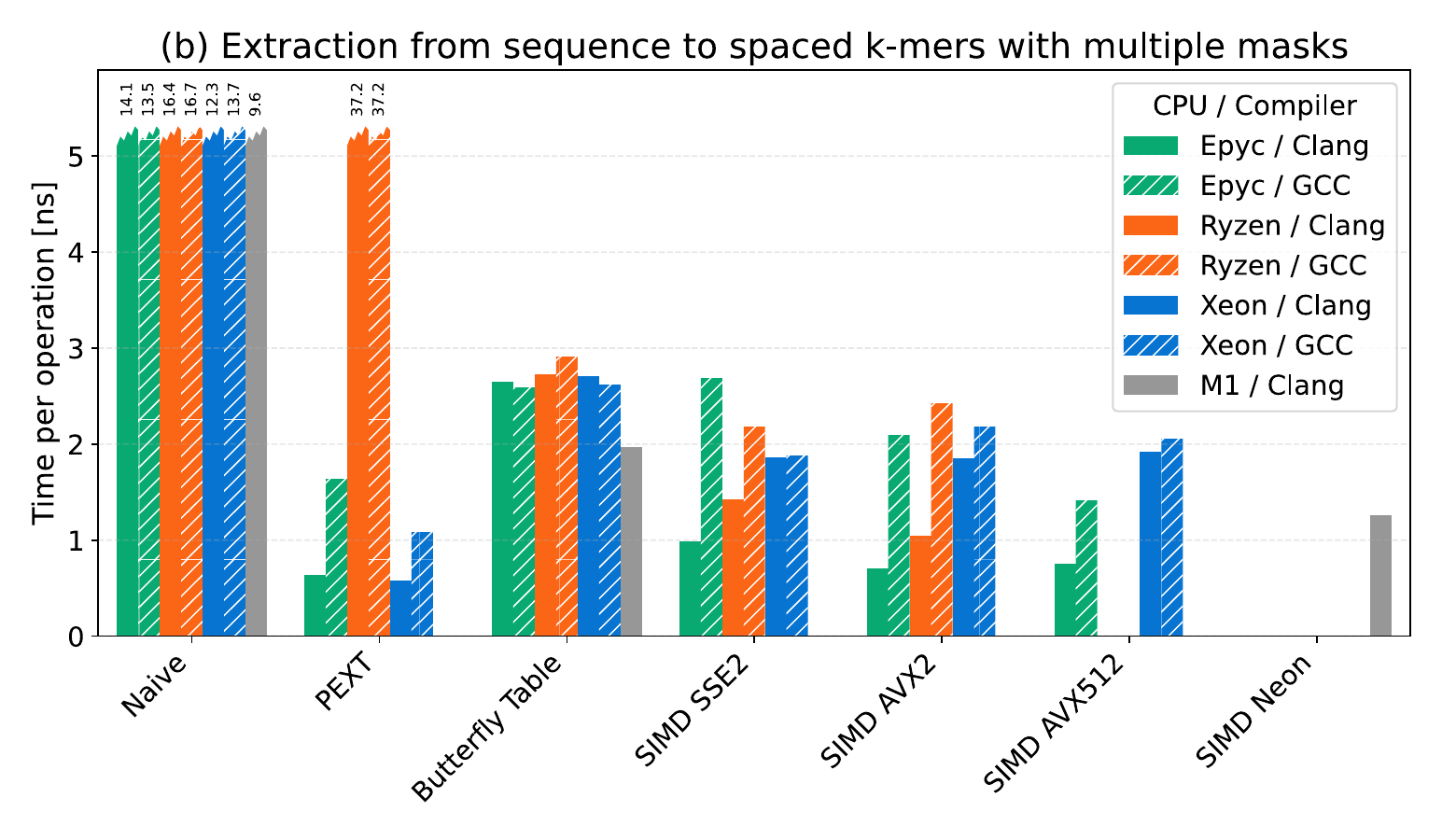}
    \begin{subfigure}{0pt}
        \phantomcaption
        \label{fig:ExtractSpacedKmers:sub:SingleMask}
    \end{subfigure}
    \begin{subfigure}{0pt}
        \phantomcaption
        \label{fig:ExtractSpacedKmers:sub:MultiMask}
    \end{subfigure}
    \caption{
        \textbf{Extracting spaced \textit{k}-mers from a sequence.}
        \textbf{(a)} Single mask. The time per $w$-mer is shown, i.\,e., total time divided by number of $w$-mers in the sequence, averaged across runs with distinct masks, for different hardware and compilers. The naive algorithm loops over the mask for each $w$-mer; the \texttt{PEXT} intrinsic is the fastest when available as a dedicated hardware instruction; the Butterfly algorithm and its SIMD-accelerated variants offer a performant alternative otherwise.
        \textbf{(b)} Multiple masks (here, 9 distinct masks). The resulting time per distinct $w$-mer is shown. Compared to a single mask, this amortizes the cost of iterating over the underlying $k$-mers in the sequence, which only needs to be performed once, followed by independent bit extractions per mask.
    }
\label{fig:ExtractSpacedKmers}
\end{figure*}


\subsection{Comparison to Existing Approaches}
\label{sec:Results:sub:Comparison}


CLARK-S \cite{Ounit2016-hi} 
is an early tool that uses spaced $k$-mers for sensitivity. It reported a $\approx$17-fold increase in runtime compared to its previous variant CLARK with non-spaced $k$-mers \cite{Ounit2015-ps}, in contrast to an expected factor of 3 due to the tool querying spaced $k$-mers with 3 distinct masks to increase sensitivity.
When inspecting the CLARK-S code, we identified several bottlenecks: It fully re-extracts $k$-mers repeatedly for each of its three masks, and uses string comparisons and branching with guaranteed branch predictor misses on the hot path (i.\,e., per character of input). Eliminating these bottlenecks lead to a 21-fold speedup in our tests on average. On the other hand, the extraction of spaced $k$-mers itself is implemented using an efficient hard-coded approach equivalent to our Block table algorithm (see Supplementary Text). In short, the slowdown of CLARK-S is not due to spaced $k$-mer extraction, but other implementation choices.

Nonetheless, the apparent slowness of CLARK-S has motivated the development of several optimized methods for spaced $k$-mer extraction, namely FSH \cite{Girotto2018-ns}, FISH \cite{Girotto2018-wt}, ISSH \cite{Petrucci2020-zt}, MISSH \cite{Mian2024-nf}, and DuoHash \cite{Gemin2026-du}. 
FSH, ISSH, and MISSH are based on the observation that typical spaced masks still do share some positions that are present in overlapping $k$-mers. They exploit the auto-correlation between these overlaps when sliding the mask along a sequence, to ``rescue'' bits that were already extracted. This however requires storing previous $w$-mers in memory during sequence iteration, and thus adds overhead. While this could be implemented with a rolling window, their current implementation seems to store all values for the whole sequence, making it not ideal to use on chromosome-length sequences.
FISH and DuoHash instead use blocks of consecutive 1s in the mask to accelerate spaced $k$-mer extraction via large lookup tables. Distinct from all other methods examined here, this allows them to compute actual $k$-mer hashes, which DuoHash showcases via the ntHash method \cite{Mohamadi2016-gg,Kazemi2022-nk}.
The overhead for overlap bookkeeping and lookup tables in these methods is however significant. Across different hardware architectures, our approaches are 3.4--4.2 times faster for single masks, and 3.2--9.5  times faster per $w$-mer when using multiple masks; details in the Supplementary Text. At the same time, our overall approach is structurally simpler to implement (with added complexity for SIMD acceleration though), and does not need to keep previous $k$-mers in memory.







Lastly, the more recent MaskJelly tool \cite{Hantze2023-uj} uses a loop over mask positions to extract each $w$-mer, thus incurring $w$ operations per $w$-mer. Implementing our approaches can at least double the performance.



\section{Discussion}
\label{sec:Discussion}

Spaced $k$-mers (i.\,e., $w$-mers) alleviate the sensitivity of $k$-mers to single mismatches, and are hence a common approach to increase robustness in sequence analyses. This comes at the expense of computational cost for extracting them from the underlying sequence, which is not as straightforward as with regular $k$-mers. Due to the gaps in the mask shifting along the sequence during traversal, the set of extracted positions in consecutive $w$-mers changes in every iteration, complicating the re-use of information between $w$-mers.

The naive approach of re-extracting each of the $w$ positions from the sequence to construct a $w$-mer is highly inefficient, in particular when combined with an inefficient encoding to transform the input sequence characters into their 2-bit representation. However, this approach unfortunately seems to be commonly used in existing tools. The previous methods FSH, FISH, ISSH, MISSH, and DuoHash improve on that baseline, but introduce significant code complexity for their data structures, and require to keep previous $k$-mers in memory.

In contrast, our approaches are simple functions that are independently applied to each $k$-mer to extract the $w$-mer. The most efficient one, the hardware \texttt{PEXT} intrinsic, only requires a $k$-mer and the mask as input. The Butterfly algorithm is the most efficient fallback software implementation; it requires an additional small precomputed table (6 $\times$ 64-bit words), and can be implemented in a few dozen lines of code. We also presented SIMD-accelerations of the latter, which introduce code complexity, but are relevant for performance-critical tools. Our dynamic selection mechanism lastly allows tools to automatically select the most performant algorithm on a given hardware.

The first introduction of what we here call the Butterfly algorithm is more than two decades old \cite{Warren2012-sy}, and the \texttt{PEXT} intrinsic more than one decade. The techniques have furthermore been known in other contexts for a while, such as in computational chess. Interestingly, despite this, neither approach seems to have been used for extracting spaced $k$-mers yet. Here, we closed this gap, and thoroughly evaluated their usage for this task. Given the data dependency between consecutive $k$-mers, and that the presented algorithms perform the minimally needed work without much overhead, we consider it difficult to make further substantial improvements.


\section{Conclusion}
\label{sec:Conclusion}

We addressed the problem of spaced $k$-mer extraction from a sequence of nucleotides for $k \leq 32$, and presented highly efficient implementations for this task, yielding a throughput of 520MB to 750MB of sequence data per second per core. They substantially outperform the commonly used naive approach by more than an order of magnitude, and advanced existing techniques by at least a factor of 3.2.
Compared to existing approaches, our algorithms are self-contained and easy to implement, and do not depend on complex data structures or previous $k$-mers along the sequence; they can hence be readily integrated into methods for processing $k$-mers.
With the presented optimizations, we remove a bottleneck for methods utilizing spaced $k$-mers. We thus conclude that the extraction of spaced $k$-mers does generally not pose a substantial barrier for high-performance computational genomics methods.





\vspace{-0.5em}
\section*{Appendices}

\subsection*{Code Availability}
The code for this project is freely available under the MIT License at \href{https://github.com/lczech/fisk}{github.com/lczech/fisk}, which contains C++ implementations of all methods and benchmarks presented.

\subsection*{Supplementary Text}
We provide details on the algorithms, their implementation, and their benchmarks in the Supplementary Text.

\subsection*{Acknowledgments}
Personal thanks to Pierre Barbera for donating compute time on his Apple M1 Pro, and to Alexandros Stamatakis and to Julian Regalado Perez for helpful feedback on this manuscript.

\subsection*{Funding}
This research project was made possible thanks to the 2023 Balzan Prize for Evolution of Humankind: Ancient DNA and Human Evolution awarded to Eske Willerslev by the International Balzan Foundation.

\subsection*{Competing Interests}
The authors declare that they have no competing interests.


\section*{References}
\bibliography{references}

\end{document}



\beginsupplement

\begingroup
\let\center\flushleft
\let\endcenter\endflushleft
\maketitle
\endgroup
\vspace*{-3.5em}

\textsuperscript{}{Section for GeoGenetics, Globe Institute, \\University of Copenhagen, Denmark.\\ Contact: \href{mailto:lucas.czech@sund.ku.dk}{lucas.czech@sund.ku.dk}}







\tableofcontents
\pagebreak


\section{Overview}
\label{supp:sec:Overview}

This supplement elaborates on the bit extract algorithms and their application to spaced $k$-mer extraction, and presents detailed benchmarks of the implementations for different hardware architectures. Our C++ implementation of all presented methods, as well as all benchmarks, are freely available under the MIT license at \href{https://github.com/lczech/fisk}{github.com/lczech/fisk}. See also our genesis C++ library \cite{Czech2020-xt} at \href{https://github.com/lczech/genesis}{github.com/lczech/genesis}, which contains additional comprehensive functionality to work with genomics data.


\subsection{Notation}
\label{sec:Overview:sub:Notation}

In this document, we use the following notation. 

\begin{itemize}
    \item $\mathrm{N} = \left\{ \text{A}, \text{C}, \text{G}, \text{T} \right\}$ is the nucleotide alphabet.
    \item $r \in \mathrm{N}^n$ is an input sequence of length $n$ over the alphabet $\mathrm{N}$, encoded as ASCII characters.
    \item A $k$-mer $\in \mathrm{N}^k$ is a sub-sequence of $r$ of length $k$, for which we use 2-bit encoding, \\ where $\text{A} = 00, \text{C} = 01, \text{G} = 10, \text{T} = 11$.
    \item $m \in \{0,1\}^k$ is a binary mask of length $k$ to select positions of an underlying $k$-mer (1s), while skipping others (0s). We call $k$ the span of the mask, and $w$ its weight, with $w = \sum_i m_i$ the number of positions in the mask that are $1$. 
    \item A $w$-mer $\in \mathrm{N}^w$ is a spaced $k$-mer, i.\,e., a $k$-mer from which positions have been extracted according to $m$.
\end{itemize}

Our goal then is to extract all consecutive $w$-mers while traversing the input sequence $r$. Typically, applications will then use those $w$-mers to, e.\,g., perform sequence searches or other relevant tasks. Here, we focus only on the extraction itself.



\subsection{Benchmarks}
\label{sec:Overview:sub:Benchmarks}

\paragraph{Input.}
Benchmarks of the bit extraction functionality were run on randomly generated 64-bit values and masks. Benchmarks of the (spaced) $k$-mer processing were run on randomly generated sequences of one million nucleotides, with a fraction of 0.001 of them randomly changed into invalid characters not in $\mathrm{N}$, which can occur in real-world data. Each benchmark was run in at least 4 repetitions, to reduce measurement noise. 

\paragraph{Data independence.}
The runtime of most algorithms presented here does not depend on the nucleotides in the sequence; hence, using randomly generated sequences is valid here for performance benchmarking. The only exception is the nucleotide encoding via a series of \texttt{if} statements, as explained below. There, performance might benefit slightly from homopolymer runs, which are unlikely in randomly generated data. As this encoding is the least performant though, and not recommended in practice, we did not further test it with real data.

\paragraph{Masks.}
The masks are the same as those used in the MISSH approach \cite{Mian2024-nf}. For single-mask benchmarks, we used a set of 9 masks with $k=31$ and $w=22$, marked as ``default'' in MISSH. For multi-mask benchmarks, each benchmark used a set of 9 masks, all applied at once to the same sequence, i.\,e., extracting 9 $w$-mers per $k$-mer. We used 5 such sets of masks, of increasing weight, with $(k,w) \in \{ (15,10), (31,14), (31, 18), (31,22), (31,26) \}$. We left out one set of masks of MISSH with $k=45$, as our approach is currently limited to $k \leq 32$.



\subsection{Hardware for Benchmarks}
\label{sec:Overview:sub:Hardware}

We benchmarked the code with the following CPUs and compilers:

\begin{itemize}
    \item AMD Epyc 7763, 64 cores, Zen 3 x86-64, released Q1/2021, tested via GitHub Actions with Clang 18.1.3 and GCC 13.3.0.
    \item AMD Epyz 9684X, 96 cores, Zen 4 x86-64, released Q2/2023, tested with Clang 17.0.0 and GCC 15.2.0.
    \item AMD Ryzen 7 PRO 4750U, 8 cores, Zen 2 x86-64, released Q2/2020, tested with Clang 17.0.6 and GCC 14.2.0.
    \item Apple M1, 8 cores, ARM, released Q4/2020, tested virtually via GitHub Actions with Apple Clang 17.0.0.
    \item Apple M1 Pro, 8 cores, ARM, released Q4/2020, tested with Apple Clang 17.0.0.
    \item Intel Xeon Platinum 8568Y+, 48 cores, x86-64, released Q4/2023, tested with Clang 17.0.0 and GCC 15.2.0.
\end{itemize}

These cover all relevant CPU architectures: AMD Epyc and Intel Xeon offer the fast hardware \texttt{PEXT} instruction (see below) for bit extraction; AMD Ryzen also has the instruction, but in slow microcode; Apple M1 does not have the instruction. These CPUs also offer distinct intrinsics, SSE2, AVX2, AVX512, and Neon, allowing us to benchmark all of them.

In this document, we only show the most relevant and distinct of those benchmarks; see \href{https://github.com/lczech/fisk}{github.com/lczech/fisk} for the full set of plots. In particular, we are leaving out the two CPUs tested via GitHub actions, and mostly focus on Clang, which has shown better performance in most of our benchmarks.



\subsection{Existing Methods and Maximum Value of k}
\label{sec:Overview:sub:ExistingMethods}

\paragraph{Limitations.}
Our approach is currently implemented for $k \leq 32$, as this allows efficient computation and storage of a $k$-mer in a 64-bit machine word, while covering the most common sizes of spaced $k$-mers. Note that this affects the initial (unspaced) $k$-mer as well as the resulting (shorter) $w$-mer, i.\,e., $w \leq k \leq 32$.
Larger values of $k$ are possible with additional bit operations or adaptation to, e.\,g., 128-bit registers, but left as future work.

\paragraph{Existing methods.}
This is mostly in line with existing work on spaced $k$-mers. 
For instance, Břinda \textit{et al.} \cite{Brinda2015-jl} investigate how spaced $k$-mers improve sensitivity in metagenomic classification; they mostly use $w \leq 22$, and only one larger mask pattern with $w=24$ and $k=33$, noting ``a drop in sensitivity''.
Similarly, Hahn \textit{et al.} \cite{Hahn2016-gw} developed a method to construct masks that maximize sensitivity for classification tasks; their approach only evaluates masks with $k < 32$.
Lastly, the Comin \& Pizzi family of methods for extracting spaced $k$-mers, namely FSH \cite{Girotto2018-ns}, FISH \cite{Girotto2018-wt}, ISSH \cite{Petrucci2020-zt}, MISSH \cite{Mian2024-nf}, and DuoHash \cite{Gemin2026-du}, evaluate 6 distinct sizes of masks, of which only the largest one has $w=32$ and $k=45$, while all others use $k \leq 32$; see our evaluation of these tools below for details.

\paragraph{Applications.}
Furthermore, tools and methods that utilize spaced $k$-mers in bioinformatics tasks also mostly use $k \leq 32$, as shown in \tabref{tab:KmerMethods}. In particular, tasks such as homology search, metagenomic classification, and short read mapping are using $k \leq 31$ and $w \leq 22$. Exceptions are tools for other purposes, where larger sequence contexts can be helpful, such as phylogenetic or pangenomic approaches.

\begin{table}[h]
    \centering
    \begin{tabular}{lll}
        Tool / Family & Purpose & Values for $k$ and $w$ \\ \hline
        PatternHunter \cite{Ma2002-hv} & Sequence homology search & $k = 18$, $w = 11$ \\
        PatternHunter 2 \cite{Li2004-qm} & Sequence homology search & $k \leq 22$, $w \leq 12$ \\
        MegaBLAST \cite{Morgulis2008-bn,Zhang2000-vp} & Sequence search & $k \leq 21$, $w \leq 12$ \\
        CLARK-S \cite{Ounit2016-hi} & Short-read taxonomic classification & $k = 31$, $w = 22$ \\
        SHRiMP2 \cite{David2011-fn} & Short-read mapping & $k \leq 20$, $w \leq 16$ \\
        BFAST \cite{Homer2009-qz} & Short-read alignment & $k \leq 40$, $w = 22$ \\
        App-SpaM \cite{Blanke2021-ah} & Phylogenetic placement & $k = 44$, $w = 12$ \\
        MaskedPanGenie \cite{Hantze2023-uj} & Pangenome genotyping & $k = 51$, $w = 31$ \\
    \end{tabular}
    \caption{\textbf{Existing bioinformatics tools that use spaced $k$-mers.} The table shows commonly used tools, along with the range of values for $k$ and $w$ that these tools use. With few exceptions, tools intended for short reads utilize $k \leq 32$.}
    \label{tab:KmerMethods}
\end{table}



\section{Encoding and Iteration of \textit{k}-mers}
\label{sec:EncodingIteration}


We encode $k$-mers using 2-bit encoding in a 64-bit word such that: 
(i) The least significant (right-most) $2k$ bits are used, while the remaining most significant (left-most) $64 - 2k$ bits are zero; 
(ii) Bit order is the same as in the input sequence: The most significant used bits represent the first character of the $k$-mer, and the least significant bits its last character.
This allows us to use regular integer comparison to get the lexicographical ordering of $k$-mers; this is not relevant here, but can be useful in other contexts. Note that not all existing implementations follow this convention.

For example, using the binary encoding \(\text{A}=00\), \(\text{C}=01\), \(\text{G}=10\), and
\(\text{T}=11\), the $k$-mer \(\texttt{GATACAGCAT}\), with \(k=10\), is encoded
as
\[
\begin{array}{c|cccccccccc}
\text{base} & \text{G} & \text{A} & \text{T} & \text{A} & \text{C} & \text{A} & \text{G} & \text{C} & \text{A} & \text{T} \\
\hline
\text{bits} & 10 & 00 & 11 & 00 & 01 & 00 & 10 & 01 & 00 & 11
\end{array}
\]

and hence occupies the least significant \(2k=20\) bits of the word as

\[
\underbrace{\mathtt{00\,\cdots\,00}}_{\mathclap{44\text{ leading zero bits}}}
\mathtt{~10~00~11~00~01~00~10~01~00~11}
\]

when storing it in a 64-bit machine word.


\subsection{Nucleotide Encoding}
\label{sec:EncodingIteration:sub:Encoding}


First, we need a function $enc(c)$ that produces the 2-bit encoding for an input character $c \in \mathrm{N}$ from ASCII encoding, or signals an invalid character if $c \notin \mathrm{N}$. To this end, in the following, in addition to the 2-bit binary values, we will also represent them by their numerical values 0-3 for the valid characters, and by value 4 as a sentinel for invalid characters.

\paragraph{Explicit checking.}
The most simplistic approach for this encoding is a \texttt{switch} statement or series of \texttt{if} statements, as shown in \algoref{alg:EncIf}, taking a character and returning its 2-bit code, applied to each character of input. For a (more or less) random input sequence, this has on average 1.5 false conditional checks, inducing a high amount of CPU pipeline flushing due to failed branch predictions. This is highly inefficient, but unfortunately often used in practice. 

\begin{algorithm}
\caption{Explicit checking using conditional statements.}
\label{alg:EncIf}
\begin{algorithmic}[1]
\Function{enc-if}{$c$} \Comment{Take a character $c$ in ASCII encoding}
    \State $c \gets c ~\mathbin{\mathrm{OR}}~ \texttt{0x20}$ \Comment{Convert ASCII letter to lower case, branchless}
    \If{$c = \texttt{'a'}$}
        \State \Return $0$
    \ElsIf{$c = \texttt{'c'}$}
        \State \Return $1$
    \ElsIf{$c = \texttt{'g'}$}
        \State \Return $2$
    \ElsIf{$c = \texttt{'t'}$}
        \State \Return $3$
    \Else
        \State \Return $4$ \Comment{Invalid character sentinel}
    \EndIf
\EndFunction
\end{algorithmic}
\end{algorithm}

The \texttt{switch} statement approach takes 5.9--8.2\,ns per character on Clang, and 0.42--0.58\,ns on GCC, which seems to implement this as a lookup (jump) table. It however cannot be relied on for the compiler to emit this code as a lookup table automatically. The \texttt{if} statement chain performs similar, with 5.1--8.1\,ns per character across architectures.


\paragraph{ASCII exploit.} 
A lesser-known low-level technique is to exploit the byte representation of the characters: The second and third least-significant bits of the ASCII code of the characters in $\mathrm{N}$ (and their lower-case equivalents) can be mangled via bit operations to obtain the desired 2-bit code. This is possible completely by chance, and was recently independently discovered and utilized for bit-compressing genomic sequences \cite{Corontzos2024-kj}. 

\pagebreak

The ASCII encodings for the nucleotide characters are:

\newcommand{\usebit}[1]{\underline{#1}}
\[
\begin{array}{c|c|c|c|c|c|c}
\text{Base} &
\text{ASCII byte} &
l &
l ~\text{\texttt{>>}}~ 1 &
l ~\text{\texttt{>>}}~ 2 &
(l ~\text{\texttt{>>}}~ 1) ~\text{\texttt{xor}}~ (l ~\text{\texttt{>>}}~ 2) &
\text{Encoding}
\\
\hline
\texttt{A} &
\mathtt{0100\,0\usebit{00}1} &
\mathtt{0\usebit{00}1} &
\mathtt{0000} &
\mathtt{0000} &
\mathtt{0000} &
\mathtt{00}
\\
\texttt{C} &
\mathtt{0100\,0\usebit{01}1} &
\mathtt{0\usebit{01}1} &
\mathtt{0001} &
\mathtt{0000} &
\mathtt{0001} &
\mathtt{01}
\\
\texttt{G} &
\mathtt{0100\,0\usebit{11}1} &
\mathtt{0\usebit{11}1} &
\mathtt{0011} &
\mathtt{0001} &
\mathtt{0010} &
\mathtt{10}
\\
\texttt{T} &
\mathtt{0101\,0\usebit{10}0} &
\mathtt{0\usebit{10}0} &
\mathtt{0010} &
\mathtt{0001} &
\mathtt{0011} &
\mathtt{11}
\end{array}
\]



The exploited bits are underlined in the table (middle bits of the lower half $l$ of the byte value).
The bits are \emph{almost} the correct encoding. A single right shift ($l$ \texttt{>>} $1$) puts those into the two rightmost bits of the result. The left of those two bits is already correct (A=C=0 and G=T=1), but the right bit is not (A=T=0 and C=G=1, but we want A=G=0 and C=T=1). We \verb|xor| with the left bit (by shifting it by one more, i.\,e., $l$ \texttt{>>} $2$) to get the desired result, as that bit is 1 for G and T, and thus flips them. Fortunately, the next bit to the left is always zero for the characters in $\mathrm{N}$, so it does not affect this operation. This works for upper and lower case, as the case bit is in the higher four bits, which are ignored here anyway. Finally, all bits except the lowest two are unset, to obtain the final encoding.

Thus, the encoding can be obtained with four bit operations:
\begin{lstlisting}[style=cppcode]
    enc(c) = ((c >> 1) xor (c >> 2)) & 0b11
\end{lstlisting}




This has the slight downside that an explicit check needs to be performed to verify the characters to be in $\mathrm{N}$, which however can also be done without branching via some additional bit operations. We refer to our implementation for details.





All operations are branchless, and can be efficiently computed. This also lends itself to a SIMD-accelerated approach, which we implemented for AVX2, encoding 32 characters at once. Across the tested CPUs, applying this encoding to a sequence takes 0.14--0.52\,ns per character (the lower value is likely due to auto-vectorization, which cannot always be applied by the compiler in more complex circumstances outside of this simple test). The approach is particularly fast when input validation can be skipped; this is usually not the case for real-world data, but can be applied if input is preprocessed already, such that invalid characters cannot occur.

\paragraph{Lookup table.}
Lastly, a common technique is a simple lookup table with 256 entries, one for each possible value of the ASCII-encoded input byte, giving either the 2-bit code or the sentinel value to indicate invalid characters. This is branchless and does not need additional valid character checks. As the table only consumes 256 bytes, it is likely to live in an efficient cache level when used in the hot loop over characters, and hence is very fast.

The lookup table has the most predictable and fastest runtime, with 0.21--0.37\,ns per character, and is hence our recommended approach for single-character encoding, only slightly outpaced by the more complex SIMD-accelerated ASCII exploit in some situations.


The effects of the encoding performance on $k$-mer extraction are shown in \figref{supp:fig:kmer_extract}.


\subsection{Consecutive \textit{k}-mer Iteration}
\label{sec:EncodingIteration:sub:Iteration}

Next, we traverse the input sequence to compute consecutive $k$-mers. 

\paragraph{Re-extraction.} The naive approach is to apply the above 2-bit encoding for each $k$-mer anew, as shown in \algoref{alg:kmer-reextract}. It takes a nucleotide sequence $S$, a value for $k$, and a function $f$ to apply to each valid $k$-mer, which represents the downstream processing of the $k$-mer as needed. It assumes a character encoding function $\operatorname{enc}(c)$ as described above, returning the 2-bit encoding for a character.

\begin{algorithm}
\caption{Naive \(k\)-mer extraction by re-extraction.}
\label{alg:kmer-reextract}
\begin{algorithmic}[1]
\Function{for-each-kmer-reextract}{$S, k, f$}

    \For{$i \gets 0$ \textbf{to} $|S| - k$} \Comment{Iterate all $k$-mers in the sequence $S$}
        \State $v \gets 0$ \Comment{The $k$-mer value using 2-bit encoding}
        \State $\operatorname{q} \gets \mathrm{true}$ \Comment{Keep track of invalid characters occurrences}

        \For{$j \gets 0$ \textbf{to} $k - 1$} \Comment{Extract all $k$ characters}
            \State $c \gets \operatorname{enc}(S[i+j])$ \Comment{Encode a single character}
            \State $\operatorname{q} \gets \operatorname{q} ~\operatorname{AND}~ (c < 4)$ \Comment{Keep track if it is valid}
            \State $v \gets (v ~\text{\texttt{<<}}~ 2) ~\operatorname{OR}~ (c ~\operatorname{AND}~ \texttt{0b11})$ \Comment{Shift in the new character value}
        \EndFor

        \If{$\operatorname{q}$} \Comment{If all characters were valid...}
            \State $f(v)$ \Comment{... process the $k$-mer}
        \EndIf
    \EndFor
\EndFunction
\end{algorithmic}
\end{algorithm}

This approach is however rather inefficient, as every input character is encoded $k$ times, in a loop with $k$ iterations for each $k$-mer. The approach is also straightforward to apply to $w$-mers, by only extracting the positions as given in the mask in the inner loop, instead of all $k$ positions. This at least avoids extracting a full $k$-mer, only to then extract the $w$-mer from it, but still needs $w$ iterations per $w$-mer.

\paragraph{Shifting.} A more efficient approach is to exploit that consecutive $k$-mers share $k-1$ of their characters. Thus, almost all encodings can be re-used between iterations, by simply bit-shifting the previous value, and erasing the surplus 2 bits remaining from the previous iteration, as shown in \algoref{alg:kmer-shift}. This rolling traversal has constant speed independent of $k$, but cannot directly be used for spaced $k$-mers, as the spaces in the mask are affecting different positions of the input sequence in each iteration. Thus, the $w$-mer needs to be extracted from the $k$-mer in a subsequent step.

\begin{algorithm}
\caption{Rolling \(k\)-mer extraction by shifting.}
\label{alg:kmer-shift}
\begin{algorithmic}[1]
\Function{for-each-kmer-shift}{$S, k, f$}

    \State $m \gets 2^{2k} - 1$ \Comment{Binary mask to remove surplus high bits}
    \State $v \gets 0$ \Comment{Current rolling $k$-mer}
    \State $\ell \gets 0$ \Comment{Counter of current valid characters}

    \For{$i \gets 0$ \textbf{to} $|S| - 1$} \Comment{Iterate all $k$-mers in the sequence $S$}
        \State $c \gets \operatorname{enc}(S[i])$ \Comment{Encode a single character}
        \State $v \gets ((v ~\text{\texttt{<<}}~ 2) ~\operatorname{AND}~ m) ~\operatorname{OR}~ (c ~\operatorname{AND}~ \texttt{0b11})$ \Comment{Shift in the encoded character}
        \State $\ell \gets \begin{cases} 
            \ell + 1, & \text{if } c < 4,\\
            0, & \text{otherwise}
        \end{cases}$ \Comment{Reset the valid counter on invalid characters}

        \If{$\ell \geq k$} \Comment{If all last $k$ characters were valid...}
            \State $f(v)$ \Comment{... process the $k$-mer}
        \EndIf
    \EndFor
\EndFunction
\end{algorithmic}
\end{algorithm}

\begin{figure}[h]
    \centering
    \begin{tabular}{@{}cc@{}}
    \includegraphics[width=0.49\linewidth]{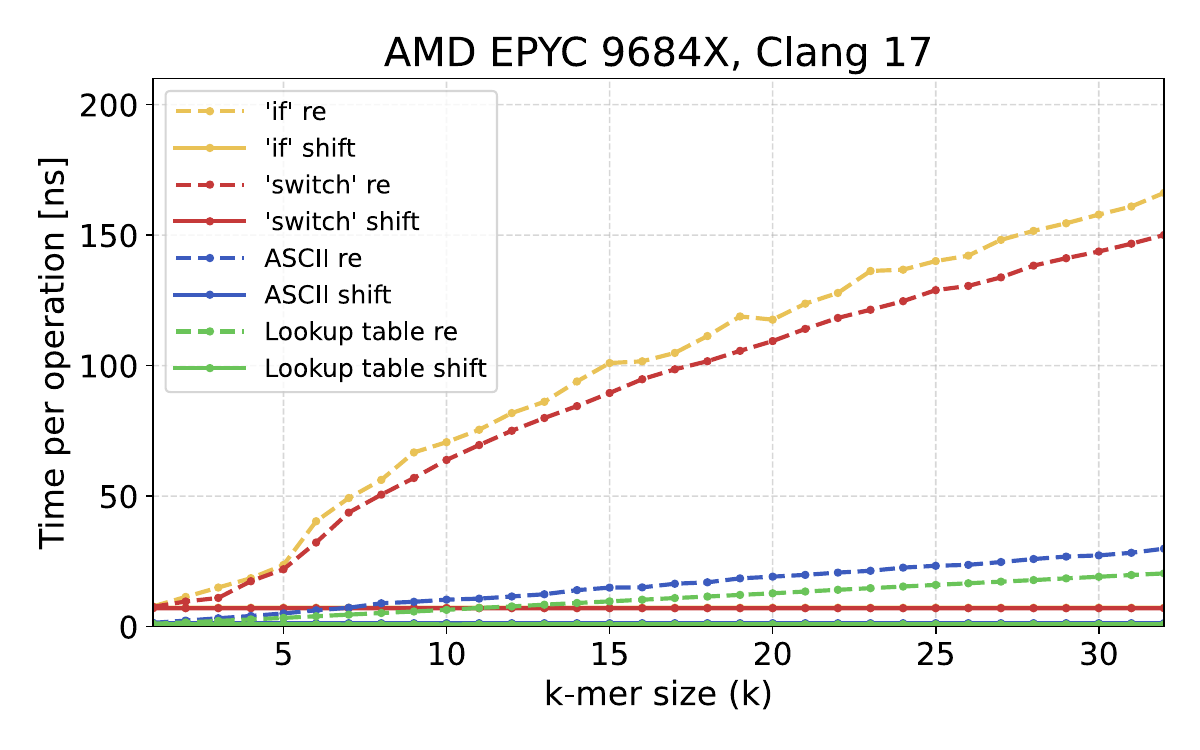} &
    \includegraphics[width=0.49\linewidth]{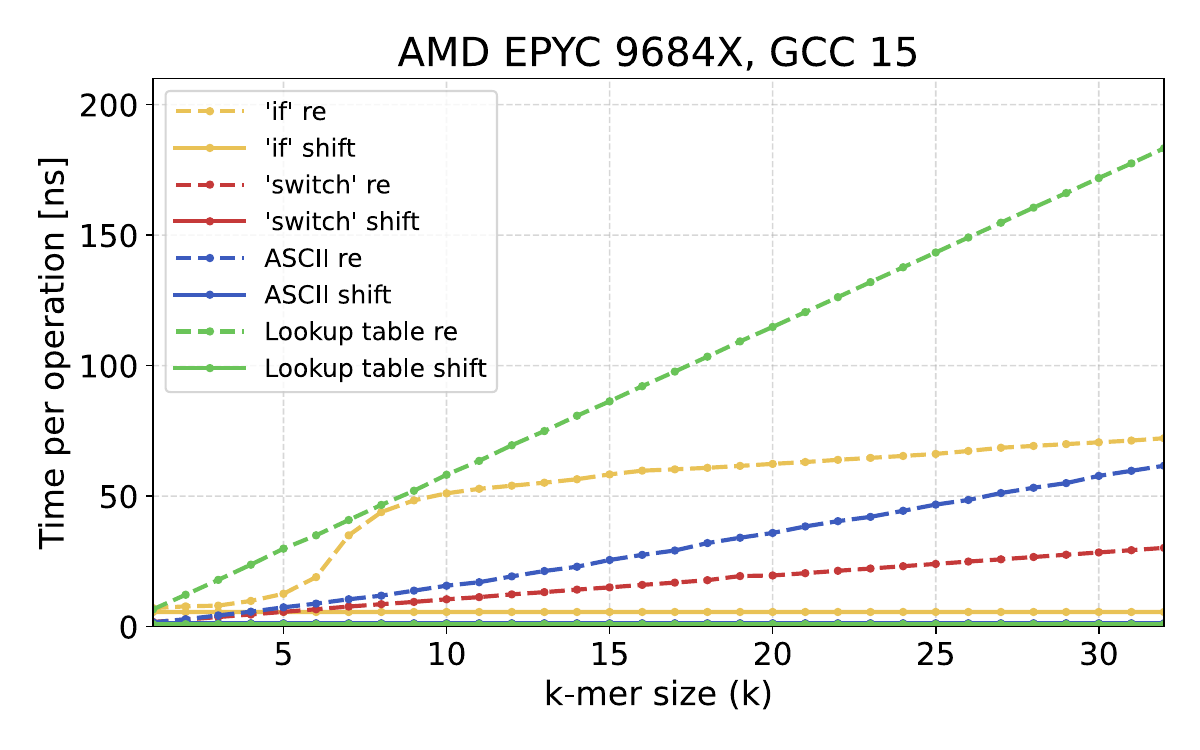}
    \\[0.5em]
    \includegraphics[width=0.49\linewidth]{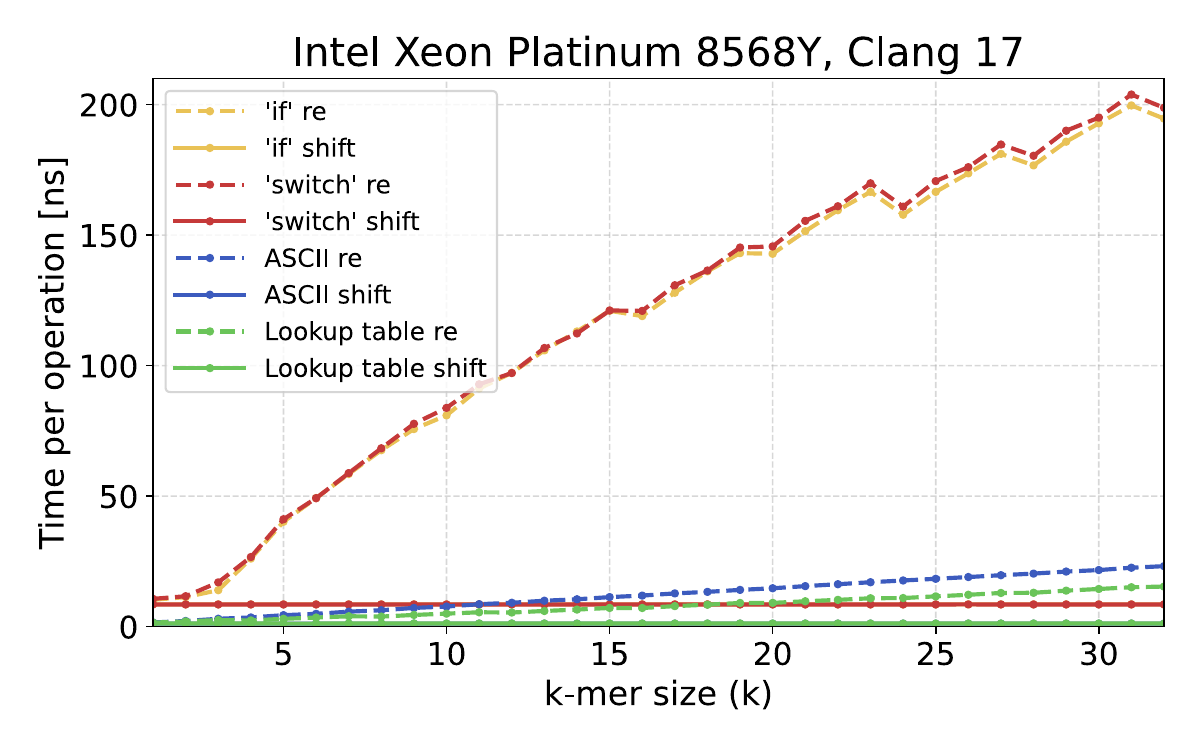} &
    \includegraphics[width=0.49\linewidth]{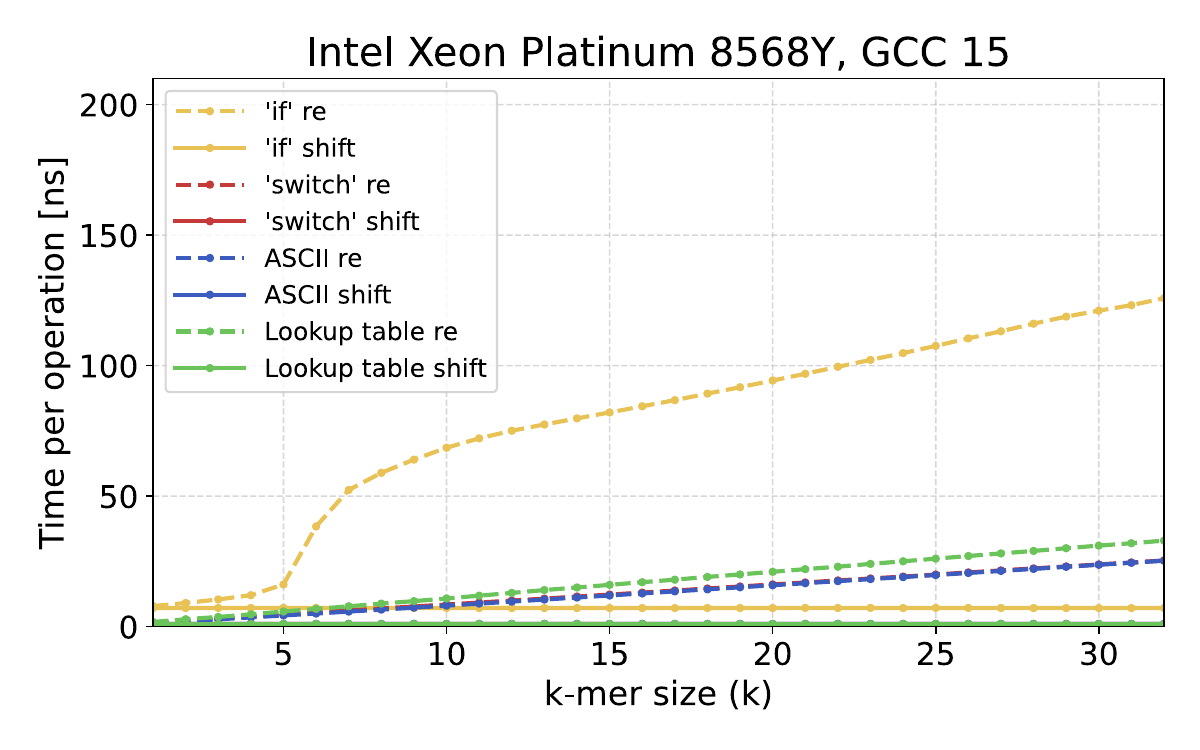}
    \end{tabular}
    \begin{subfigure}{0pt}
        \phantomcaption
        \label{fig:kmer_extract:sub:AMD_EPYC_9684X_Clang_17}
    \end{subfigure}
    \begin{subfigure}{0pt}
        \phantomcaption
        \label{fig:kmer_extract:sub:AMD_EPYC_9684X_GCC_15}
    \end{subfigure}
    \begin{subfigure}{0pt}
        \phantomcaption
        \label{fig:kmer_extract:sub:Intel_Xeon_Platinum_8568Y_Clang_17}
    \end{subfigure}
    \begin{subfigure}{0pt}
        \phantomcaption
        \label{fig:kmer_extract:sub:Intel_Xeon_Platinum_8568Y_GCC_15}
    \end{subfigure}
    \vspace*{-1em}
    \caption{
        \textbf{Extracting $k$-mers from a sequence.} For different values of $k \in [0, 32]$, time per $k$-mer on two server CPUs is shown, with different compilers. Line colors correspond to the function used for converting input ASCII characters into 2-bit encoding; line styles are dashed for the approach of extracting the $k$-mer at each position anew (``re-extract''), and solid for the shifting updates of consecutive $k$-mers between iterations.
        This shows two important observations: (i) The costs for re-extracting each $k$-mer quickly become devastating even for smaller values of $k$. (ii) Compilers behave differently in their optimizations of the encoding function, even on the same architecture.
        The most efficient way of iterating $k$-mers is hence to use the branch-less encoding function via lookup table, and shifting $k$-mer traversal, as shown in the bottom green line in all sub-figures, which is barely above the x-axis here, near the mark of 1\,ns per $k$-mer.
    }
\label{supp:fig:kmer_extract}
\end{figure}

Benchmarks for the two approaches for $k$-mer extraction, combined with the above variants of character encodings, are shown in \figref{supp:fig:kmer_extract}. For obvious reasons, the second approach is the predominant one in high-performance $k$-mer methods. We mention both, as the first approach is still occasionally used in practice, likely because it is easiest to implement, as it does not need to keep track of $k$-mers across iterations.

Note that a $k$-mer depends on all $k-1$ instances before it to be constructed, which makes this problem hard to optimize. The two outlined approaches thus suffer from different inefficiencies: Re-extraction circumvents the data dependency, but requires repeated encoding of the characters; the shifting approach avoids the repeated encoding, but instead is limited in throughput due to the data dependency. Still, by far the fastest approach is to combine the lookup table encoding with the shift-based iteration, performing at $\approx$1\,ns per character. 

\clearpage
\newpage


\section{Bit Extract}
\label{sec:BitExtract}

We now have an efficient loop over all $k$-mers, extracting (unspaced) $k$-mers sliding along the input sequence. For each $k$-mer, we then need to apply the mask $m$ and extract the positions set in $m$ to construct the desired $w$-mer. 

\paragraph{Naive approach.}
As mentioned above, one ``naive'' way to do this is to skip the construction of ther $k$-mer altogether, and instead re-extract the $w$-mer by directly encoding the characters set in the mask from the input sequence, with a loop with $w$ iterations per $w$-mer, similar to \algoref{alg:kmer-reextract}. This is the baseline which the family of methods of FSH \cite{Girotto2018-ns}, FISH \cite{Girotto2018-wt}, ISSH \cite{Petrucci2020-zt}, MISSH \cite{Mian2024-nf}, and DuoHash \cite{Gemin2026-du} improves upon. We also use this as the baseline here.



\begin{figure}[!htb]
    \centering
    \includegraphics[width=0.95\textwidth]{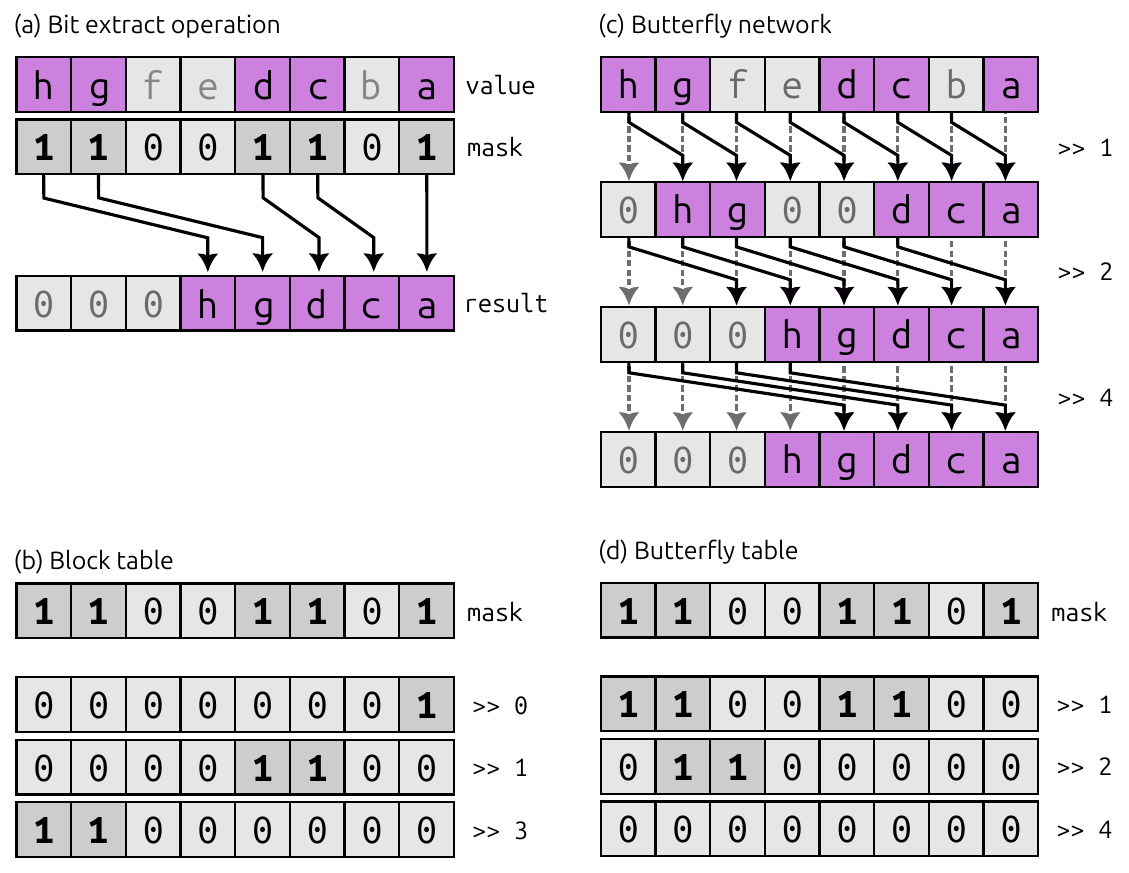}
    \begin{subfigure}{0pt}
        \phantomcaption
        \label{fig:BitExtract:sub:BitExtract}
    \end{subfigure}
    \begin{subfigure}{0pt}
        \phantomcaption
        \label{fig:BitExtract:sub:BlockTable}
    \end{subfigure}
    \begin{subfigure}{0pt}
        \phantomcaption
        \label{fig:BitExtract:sub:ButterflyNetwork}
    \end{subfigure}
    \begin{subfigure}{0pt}
        \phantomcaption
        \label{fig:BitExtract:sub:ButterflyTable}
    \end{subfigure}
    \caption{
        \textbf{Bit extraction from words.}
        \textbf{(a)} The bit extract operation. Bits from a value are ``extracted'' according to a binary mask, and transferred to the lower bits of the resulting word. In other words, the ``gaps'' caused by 0s in the mask are closed. For simplicity, we here only show 8 bits worth of data; the actual instruction applies to 64-bit words. 
        \textbf{(b)} The block table is a decomposition of the mask into blocks of consecutive 1s, accompanied by the respective amount of shifting the block needs to receive to move to the target positions. It has as many rows as blocks of 1s in the mask.
        \textbf{(c)} The butterfly network connects each input bit to each output bit right of it (solid black arrows), or to itself (dashed gray arrows), via a decomposition into shifts of increasing powers of two. This allows any input value bit to be shifted to any relevant output bit.
        \textbf{(d)} The butterfly network for a particular mask consists of a table, which indicates for each input bit which path it takes through the network, i.\,e., which of the shifts are applied in each step.
        For instance, the two bits in positions c and d need to be shifted by 1, and hence only occur in the first row of the table. The bits in positions g and h need to be shifted by $3 = 1+2$ in total; the first row of the table marks them for a shift by 1, after which row two then marks them for a shift by 2.
        For this 8-bit word, only the first three powers of two are needed; generally, for $b$-bit words, the table has $\log_2(b)$ entries, i.\,e., 6 entries for 64-bit words.
        \textbf{Note:} For generality, we show the examples here with individual bits. In our use case however, a \textit{k}-mer is typically represented with 2 bits per nucleotide. Thus, in practice, for each symbol of the spaced mask, 2 consecutive bits need to be set in the masks here to extract a nucleotide. This has no effect on efficiency.
    }
\label{supp:fig:BitExtract}
\end{figure}

\paragraph{Bit shifting approach.}
In our improved approaches, we instead process the $k$-mers as in \algoref{alg:kmer-shift}, which allows fast iteration with bit shifting, and apply a generic bit extraction function to the $k$-mer to get the $w$-mer, see \figref{fig:BitExtract:sub:BitExtract} for an example. In other words, we keep only the positions from a given $k$-mer that are set in the mask $m$, and pack them densely into the lower bits of the resulting $w$-mer, i.\,e., close the gaps left in the mask.

\paragraph{History.}
The problem of bit extraction from machine words is more general and useful in other applications. Interestingly, optimized code for the operation has been developed in the chess community\footnote{%
  \href{https://www.chessprogramming.org/BMI2}{\textcolor{black}{chessprogramming.org/BMI2}}%
}. In chess software, 64-bit words are used to succinctly represent the 64 squares on a chess board; thus, bit extraction can be used to test positions and potential moves. Compared to chess and other applications, we here usually have the advantage of having a fixed mask (or small set of masks) across the runtime of our code. That is, masks are given once initially, but then do not change throughout the program, as this would induce a different spacing pattern. We can exploit this to pre-compute a representation of the mask(s) that allows fast application to incoming values (the $k$-mers).

\paragraph{Implementation and benchmarking.}
In the following, we describe different bit extraction algorithms and their implementations; their benchmarks are shown in \figref{supp:fig:BitExtractWeights}--\ref{supp:fig:BitExtractSummary}. 






\subsection{Hardware (PEXT)}
\label{sec:BitExtract:sub:PEXT}

\urldef{\IntelPextPrinted}\nolinkurl{intel.com/content/www/us/en/docs/intrinsics-guide/index.html#text=pext&ig_expand=5088}

Modern x86 hardware supports the \texttt{PEXT} (parallel bit extract) intrinsic%
\footnote{%
  \href{https://www.intel.com/content/www/us/en/docs/intrinsics-guide/index.html\#text=pext\&ig_expand=5088}{\IntelPextPrinted}%
}
of BMI2 (Bit Manipulation Instruction Set 2). The name ``parallel'' indicates how it is implemented in hardware; it is however a single operation on a 64-bit word.

\paragraph{History.}
On Intel CPUs, it is available since the Haswell architecture (Q2/2013) with a latency of 3 cycles, and a throughput of 1 instruction per cycle. On AMD CPUs, the instruction is available since the Excavator architecutre (Q2/2015), but implemented in slow microcode, with a runtime dependent on the weight of the mask, typically 18-20 cycles at least. Only since the Zen 3 architecture (Q4/2020) have AMD CPUs implemented \texttt{PEXT} in actual hardware. On ARM hardware (such as the Apple M series of CPUs), it is not available. Our benchmarks include CPU models testing all these different architectures.

\subsection{Bit Loop}
\label{supp:sec:BitExtract:sub:BitLoop}


A first naive approach is to loop over the bits set in the mask, keeping track of their position in the mask and the positions already extracted. This is shown in \algoref{alg:bit-extract-bitloop}, for input value $x$ and mask $m$. This has runtime linearly increasing with the weight of the mask; for our spaced $k$-mer approach, it is even worse than re-extracting $w$-mers from the input sequence, as we need two bits per nucleotide here.

\begin{algorithm}
\caption{Bit extraction by iterating over selected bits.}
\label{alg:bit-extract-bitloop}
\begin{algorithmic}[1]
\Function{bit-extract-bitloop}{$x, m$}
    \State $y \gets 0$
    \State $b \gets 1$

    \While{$m \neq 0$}
        \State $\ell \gets \operatorname{lsb}(m)$
            \Comment{Least Significant Bit of \(m\)}

        \If{$(x ~\operatorname{AND}~ \ell) \neq 0$} \Comment{If the $\ell$ bit is set in the value $x$...}
            \State $y \gets y ~\operatorname{OR}~ b$ \Comment{... set the next bit in $y$}
        \EndIf

        \State $m \gets m ~\operatorname{XOR}~ \ell$ \Comment{Move to the next set bit in the mask}
        \State $b \gets b ~\text{\texttt{<<}}~ 1$
    \EndWhile

    \State \Return $y$
\EndFunction
\end{algorithmic}
\end{algorithm}

Here, \(\operatorname{lsb}(m)\) denotes the least significant set bit of \(m\),
for instance computed as \verb|(m) AND (-m)|.


\subsection{Byte Table}
\label{supp:sec:BitExtract:sub:ByteTable}


In this approach, we pre-compute a lookup table for each byte of input and byte of the mask, i.\,e., a $256 \times 256$ sized table, denoting the extracted bits for each combination, as well as a smaller table with the corresponding bit weight of the mask byte. Then, to extract a value, the extracted value of each corresponding pair of bytes of the value and mask is retrieved from the table, and shifted according to the mask bit count. For simplicity, we are not showing pseudo-code of this approach here, and refer to our implementation instead.

This approach is a reasonably fast drop-in software equivalent of \texttt{PEXT}, with constant speed independent of the mask, always using exactly 8 iterations (one per byte of the 64-bit word).
However, for a fixed mask, further speed can be gained via pre-processing of the mask.


\subsection{Block Table}
\label{supp:sec:BitExtract:sub:BlockTable}


\paragraph{Concept.}
The insight we exploit in this approach is that typical masks for spaced $k$-mers have longer runs (blocks) of consecutive 1s in the mask (all tested masks have at most 8 such blocks). We can thus pre-process the mask into a table containing a row for each block, keeping track of a mask to extract that block (an \verb|and| operation) and a value to shift (a \texttt{>}\texttt{>} operation) it to the desired position in the output. Applying this to an input word then is a simple loop over the rows, performing both operations for each row, as shown in \figref{fig:BitExtract:sub:BlockTable}.

\paragraph{Preprocessing.}
We show the pre-processing in \algoref{alg:bit-extract-block-table-preprocess}, which takes a mask $m$ and returns the block table $T$. For each maximal run of consecutive 1-bits in the mask \(m\), the pre-processing stores a row in the table, consisting of a pair \((r,d)\), where \(r\) selects the run at its original position and \(d\) is the right-shift needed to move the selected bits into their packed output position. 
At most, this needs 32 active rows in the table, which happens with an alternating pattern of 1s and 0s in the mask. This can be implemented with a fixed-sized table. To indicate the end of the needed (active) rows, a final $(0,0)$ sentinel value is appended to $T$, which eliminates the need for an additional variable to store the length of the table (which is equal to the number of blocks of consecutive 1-bits, and hence depends on the mask $m$).


\begin{algorithm}
\caption{Preprocessing a block table for bit extraction.}
\label{alg:bit-extract-block-table-preprocess}
\begin{algorithmic}[1]
\Function{preprocess-block-table}{$m$}
    \State $T \gets$ empty table
    \State $b \gets 0$
        \Comment{Current input bit position}
    \State $p \gets 0$
        \Comment{Next output bit position}

    \While{$b < 64$}
        \If{$((m ~\text{\texttt{>>}}~ b) ~\operatorname{AND}~ 1) = 0$} \Comment{Skip zero bits}
            \State $b \gets b + 1$
            \State \textbf{continue}
        \EndIf

        \State $s \gets b$
            \Comment{Start of a run of ones}

        \While{$b < 64$ \textbf{and} $((m ~\text{\texttt{>>}}~ b) ~\operatorname{AND}~ 1) = 1$} \Comment{Find the last set bit in the run}
            \State $b \gets b + 1$
        \EndWhile

        \State $\ell \gets b - s$
            \Comment{Length of the run}
        \State $r \gets ((1 ~\text{\texttt{<<}}~ \ell) - 1) ~\text{\texttt{<<}}~ s$
            \Comment{Mask of consecutive bits for selecting this run}
        \State $d \gets s - p$
            \Comment{Right shift needed to pack into output position}

        \State append $(r,d)$ to $T$
        \State $p \gets p + \ell$ \Comment{Increment next output bit by length of run}
    \EndWhile

    \State append $(0,0)$ to $T$
        \Comment{Sentinel to mark end of table}
    \State \Return $T$
\EndFunction
\end{algorithmic}
\end{algorithm}

\paragraph{Application.}
Once the table is computed, the bit extraction is shown in \algoref{alg:bit-extract-block-table}, taking an input value $x$ and the table $T$, and returning the extracted value. It applies all rows of $T$ independently and combines them via an \verb|or| operation, and stops at the sentinel value. As each row is independent, it is well suited for CPU instruction pipelining, which can be further increased through loop unrolling. For instance, with a typical spaced $k$-mer mask with 8 blocks of consecutive 1s, extraction with 8-fold unrolling results in a single loop iteration.


\begin{algorithm}
\caption{Bit extraction using a precomputed block table.}
\label{alg:bit-extract-block-table}
\begin{algorithmic}[1]
\Function{bit-extract-block-table}{$x, T$}
    \State $y \gets 0$ \Comment{Extracted value}
    \State $i \gets 0$ \Comment{Table row counter}

    \While{$T[i] \neq (0,0)$} \Comment{Iterate all entries until the sentinel}
        \State $(r,d) \gets T[i]$ \Comment{Entry in the table}
        \State $y \gets y ~\operatorname{OR}~ ((x ~\operatorname{AND}~ r) ~\text{\texttt{>>}}~ d)$ \Comment{Extract the block, and shift into position}
        \State $i \gets i + 1$
    \EndWhile

    \State \Return $y$
\EndFunction
\end{algorithmic}
\end{algorithm}

\paragraph{Comparison.}
The approach is used by CLARK-S \cite{Ounit2016-hi}, albeit for a set of three hard-coded masks only. This allows them to operate without any loop, eliminating a conditional brach on the hot path. Our implementation allows arbitrary masks.
The idea is furthermore similar to the FISH method \cite{Girotto2018-wt}, in that it exploits the block structure of consecutive 1s in the mask. However, FISH constructs a data structure with distinct lookup tables separated by the run lengths of blocks, which introduces overhead for pointer chasing in those tables.






\subsection{Butterfly Network}
\label{supp:sec:BitExtract:sub:ButterflyNetwork}


\paragraph{Concept.}
The most efficient approach in our tests is to pre-processes the mask as follows: For every bit set (1) in the mask, we get the number of unset (0) bits right of it, which is the amount by which the corresponding input value bit at this position needs to be shifted to its target position towards the lower end of the word. This amount is decomposed into its sum of powers of two, such that the respective shifts can be consecutively applied. This way, any input bit can be right-shifted to any target output bit, as shown in \figref{fig:BitExtract:sub:ButterflyNetwork}. We call this the Butterfly network, inspired by similar bit-mixing techniques from the literature \cite{Weinstein1969-pk}; admittedly, our right-shifting network only resembles one wing of the butterfly. The technique was originally described in the book ``Hacker's Delight'' \cite{Warren2012-sy}, but remained unnamed therein.

\paragraph{Preprocessing.}
For a given mask, the network is applied to an input value in a series of steps, applying the powers-of-two shifts to each bit as needed. The steps are pre-processed into rows of a table, indicating which bits need to be shifted at each step; see \figref{fig:BitExtract:sub:ButterflyTable} for an example. For $b$-bit words, the table has $\log_2(b)$ entries, each of size $b$ bits itself, i.\,e., 6 entries for 64-bit words, corresponding to shifts 1, 2, 4, 8, 16, 32.
The pre-processing is shown in \algoref{alg:bit-extract-butterfly-preprocess}, for a given mask $m$. We refer to the original source \cite{Warren2012-sy} for an alternative explanation of the algorithm and involved bit operations.
Note that instead of pre-computing the step values, they can be computed in parallel with their application to the input, at the cost of additional bit operations. This variant of the algorithm is relevant for generic use cases without a fixed set of masks, but omitted here for simplicity. 




\begin{algorithm}
\caption{Preprocessing a butterfly table for bit extraction.}
\label{alg:bit-extract-butterfly-preprocess}
\begin{algorithmic}[1]
\Function{preprocess-butterfly-table}{$m$}
    \State $M \gets m$
    \State $Z \gets (\operatorname{NOT}~ M) ~\text{\texttt{<<}}~ 1$
        \Comment{Candidate positions}

    \For{$i \gets 0$ \textbf{to} $5$} \Comment{$\log_2(64) = 6$ iterations are needed}

        \State $P \gets Z ~\operatorname{XOR}~ (Z ~\text{\texttt{<<}}~ 1)$ \Comment{Decompose candidate bits into powers of two}
        \State $P \gets P ~\operatorname{XOR}~ (P ~\text{\texttt{<<}}~ 2)$
        \State $P \gets P ~\operatorname{XOR}~ (P ~\text{\texttt{<<}}~ 4)$
        \State $P \gets P ~\operatorname{XOR}~ (P ~\text{\texttt{<<}}~ 8)$
        \State $P \gets P ~\operatorname{XOR}~ (P ~\text{\texttt{<<}}~ 16)$
        \State $P \gets P ~\operatorname{XOR}~ (P ~\text{\texttt{<<}}~ 32)$

        \State $s \gets 2^i$
            \Comment{Current shift distance (power of two)}
        \State $V \gets P ~\operatorname{AND}~ M$
            \Comment{Bits to move by \(s\) positions}
        \State $T[i] \gets V$ \Comment{Store in current table row}

        \State $M \gets (M ~\operatorname{XOR}~ V) ~\operatorname{OR}~ (V ~\text{\texttt{>>}}~ s)$
            \Comment{Apply this stage to the mask}
        \State $Z \gets Z ~\operatorname{AND}~ (\operatorname{NOT}~ P)$ \Comment{Update the candidate position}
    \EndFor

    \State \Return $T$
\EndFunction
\end{algorithmic}
\end{algorithm}

\begin{figure}[!htb]
    \centering
    \begin{tabular}{@{}cc@{}}
    \includegraphics[width=0.49\linewidth]{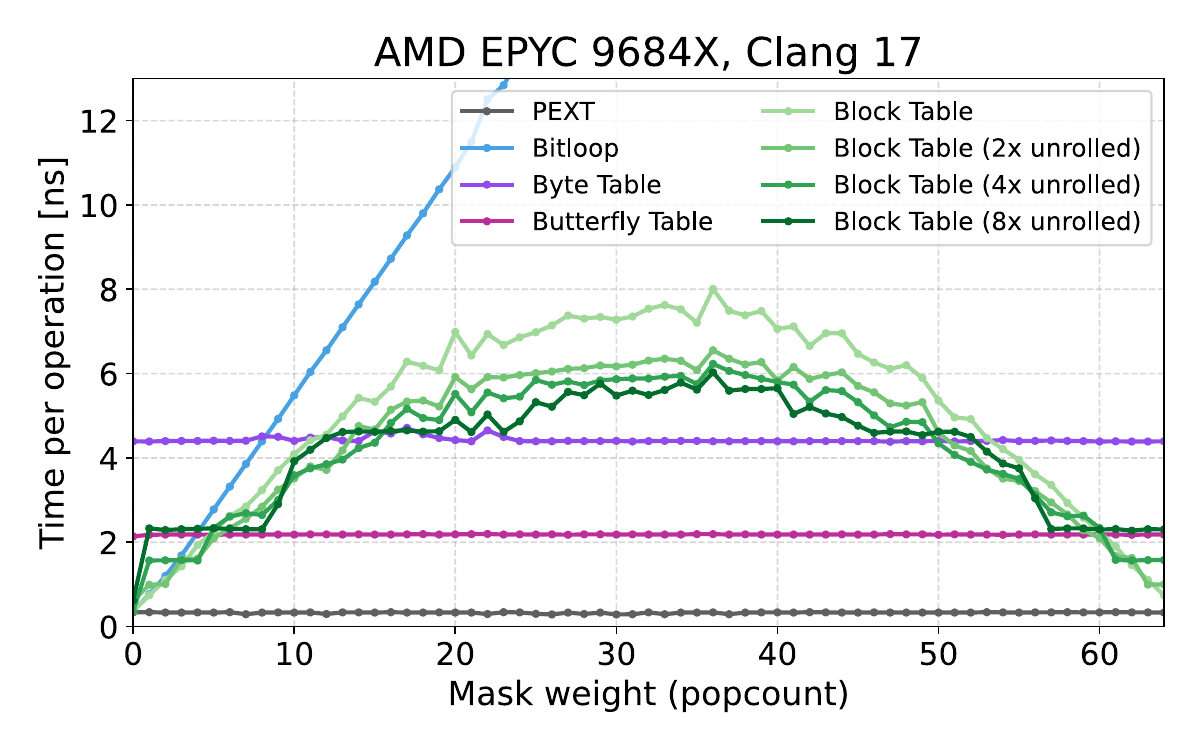} &
    \includegraphics[width=0.49\linewidth]{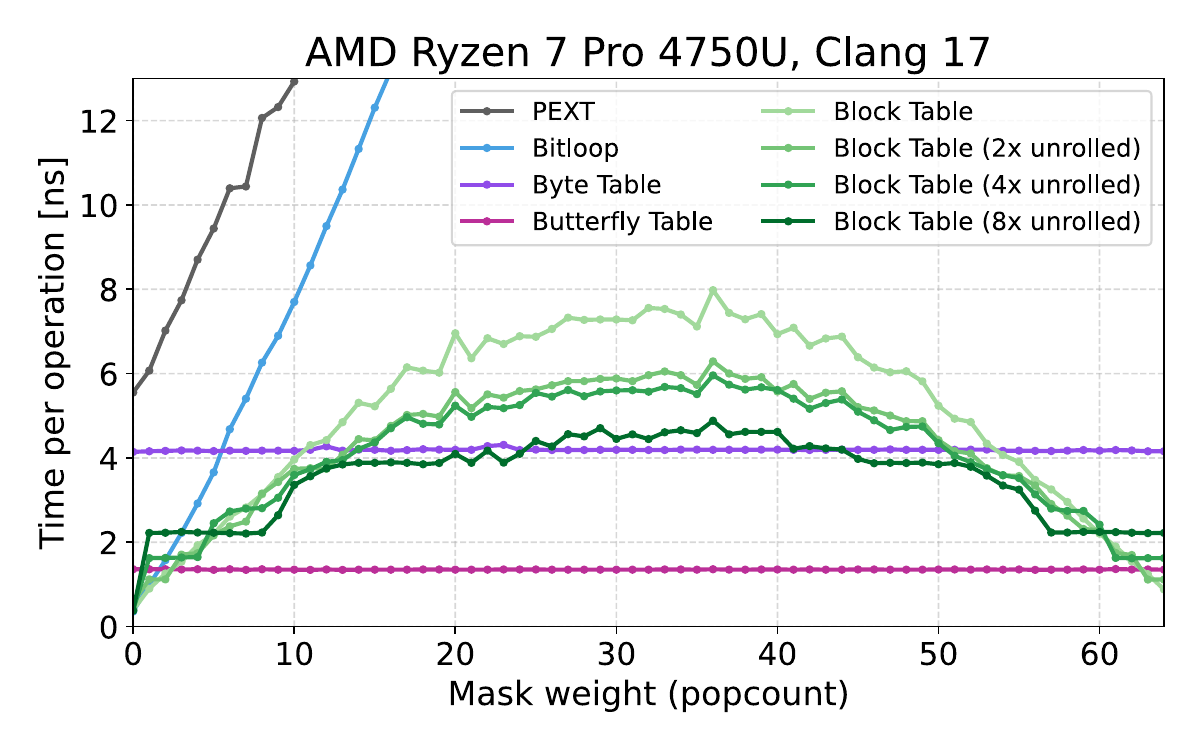}
    \\[0.5em]
    \includegraphics[width=0.49\linewidth]{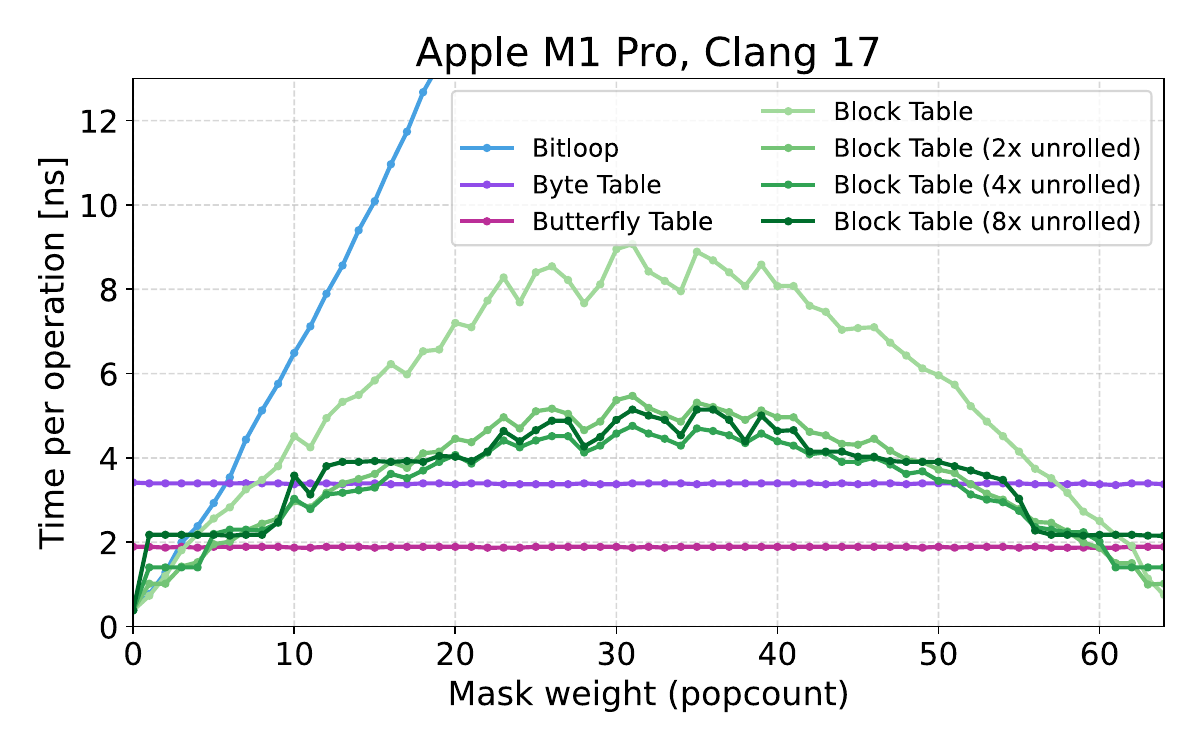} &
    \includegraphics[width=0.49\linewidth]{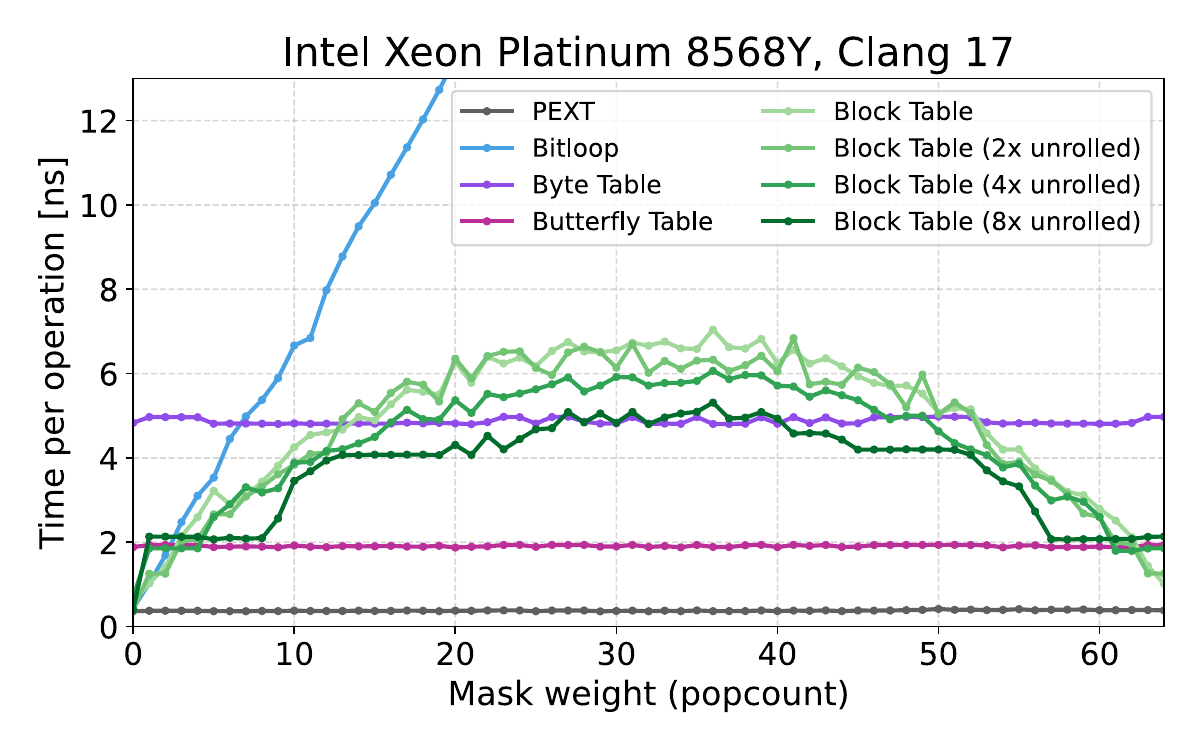}
    \end{tabular}
    \begin{subfigure}{0pt}
        \phantomcaption
        \label{fig:BitExtractWeights:sub:AMDEpyc9684XClang17}
    \end{subfigure}
    \begin{subfigure}{0pt}
        \phantomcaption
        \label{fig:BitExtractWeights:sub:AMDRyzen7Pro4750UClang17}
    \end{subfigure}
    \begin{subfigure}{0pt}
        \phantomcaption
        \label{fig:BitExtractWeights:sub:AppleM1ProClang17}
    \end{subfigure}
    \begin{subfigure}{0pt}
        \phantomcaption
        \label{fig:BitExtractWeights:sub:IntelXeonPlatinum8568YClang17}
    \end{subfigure}
    \vspace*{-1em}
    \caption{
        \textbf{Bit extraction performance for different weights (number of set bits) of the mask.}
        The hardware \texttt{PEXT} is by far the most performant where available as a dedicated intrinsic; note however that on the older AMD Ryzen 7, where it is implemented in CPU microcode instead, performance is dependent on the mask weight (number of bits set in the mask), and performs worse even than our naive bitloop implementation in software. The block table implementation has as many steps as runs of consecutive 1s in the mask; for the randomly generated inputs used here, this thus has a bell shape. Generally, this implementation benefits from loop unrolling.
        The most consistently fast software implementation is however the Butterfly network, with performance independent of the mask weight.
    }
\label{supp:fig:BitExtractWeights}
\end{figure}

\begin{figure}[!htb]
    \centering
    \begin{tabular}{@{}cc@{}}
    \includegraphics[width=0.49\linewidth]{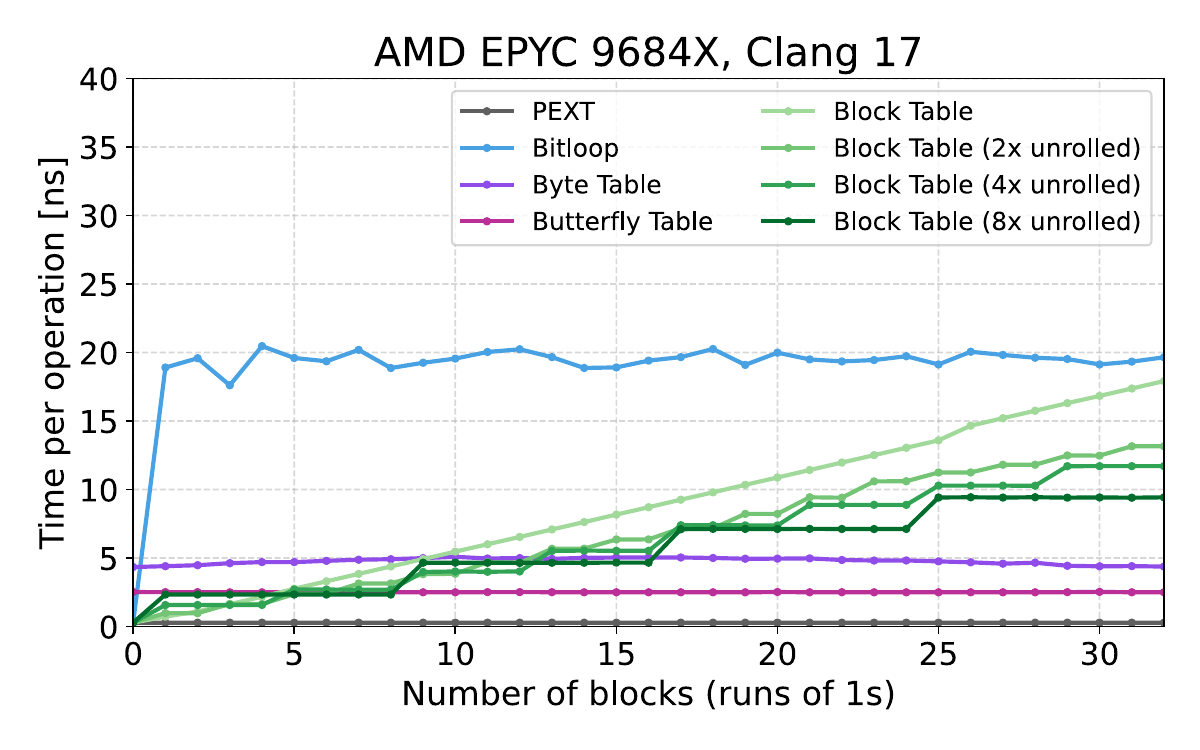} &
    \includegraphics[width=0.49\linewidth]{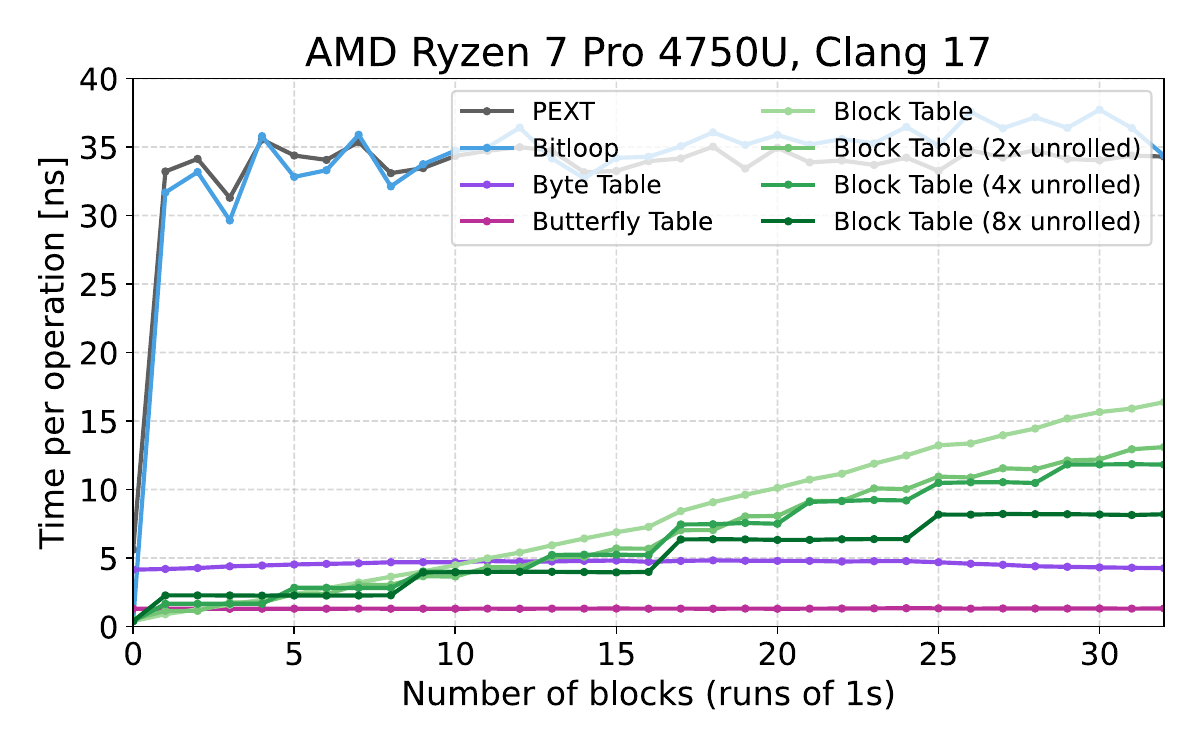}
    \\[0.5em]
    \includegraphics[width=0.49\linewidth]{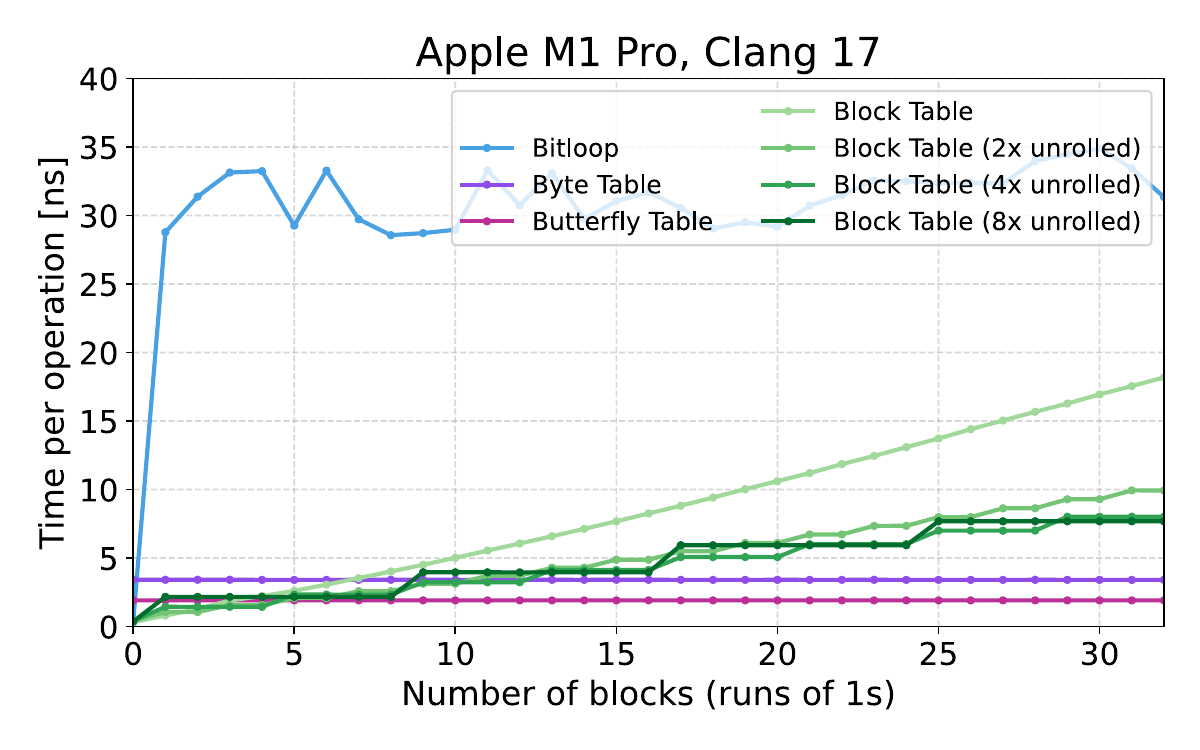} &
    \includegraphics[width=0.49\linewidth]{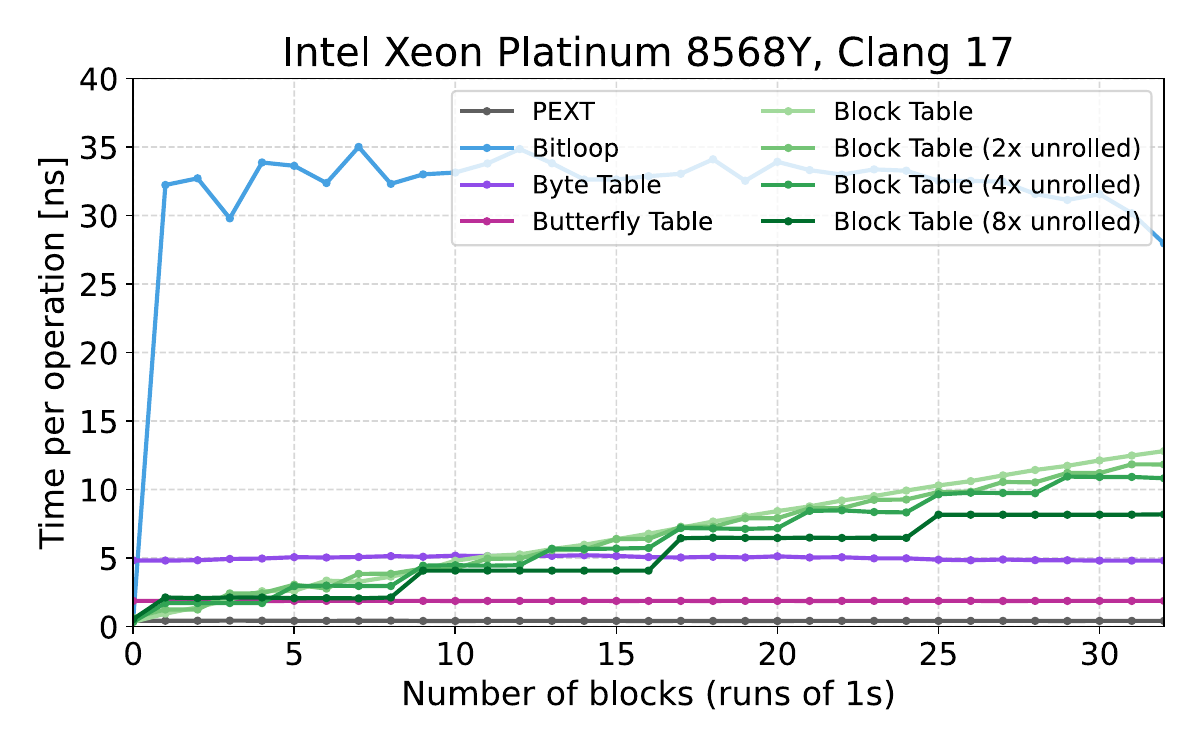}
    \end{tabular}
    \begin{subfigure}{0pt}
        \phantomcaption
        \label{fig:BitExtractBlocks:sub:AMDEpyc9684XClang17}
    \end{subfigure}
    \begin{subfigure}{0pt}
        \phantomcaption
        \label{fig:BitExtractBlocks:sub:AMDRyzen7Pro4750UClang17}
    \end{subfigure}
    \begin{subfigure}{0pt}
        \phantomcaption
        \label{fig:BitExtractBlocks:sub:AppleM1ProClang17}
    \end{subfigure}
    \begin{subfigure}{0pt}
        \phantomcaption
        \label{fig:BitExtractBlocks:sub:IntelXeonPlatinum8568YClang17}
    \end{subfigure}
    \vspace*{-1em}
    \caption{
        \textbf{Bit extraction performance for different number of runs of consecutive 1s (blocks) in the mask.}
        Here, instead of increasing the weight of the mask, we tested different number of runs of consecutive 1s, each with a random length (within the constraints of a 64-bit word). The maximally possible number of runs is 32, corresponding to an alternating pattern of 1s and 0s. The bit extract implementations whose runtime depends on the weight of the mask (bitloop and \texttt{PEXT} on the older AMD Ryzen 7) hence exhibit a similar runtime across masks, as the number of bits set is on average half the mask. The effect of loop unrolling in the block table approach is clearly visible here in form of ``steps'' at each level of unrolling; with 8-fold unrolling, the approach is competitive with the butterfly network for the runs of 1s that are most relevant for spaced $k$-mers (up to 8 blocks). Again though, across all masks, the fast hardware \texttt{PEXT} as well as the butterfly network implementation dominate performance here, and are thus the two preferred approaches.
    }
\label{supp:fig:BitExtractBlocks}
\end{figure}

\begin{figure}[!htb]
    \centering
    \begin{tabular}{@{}c@{}}
    \includegraphics[width=0.8\linewidth]{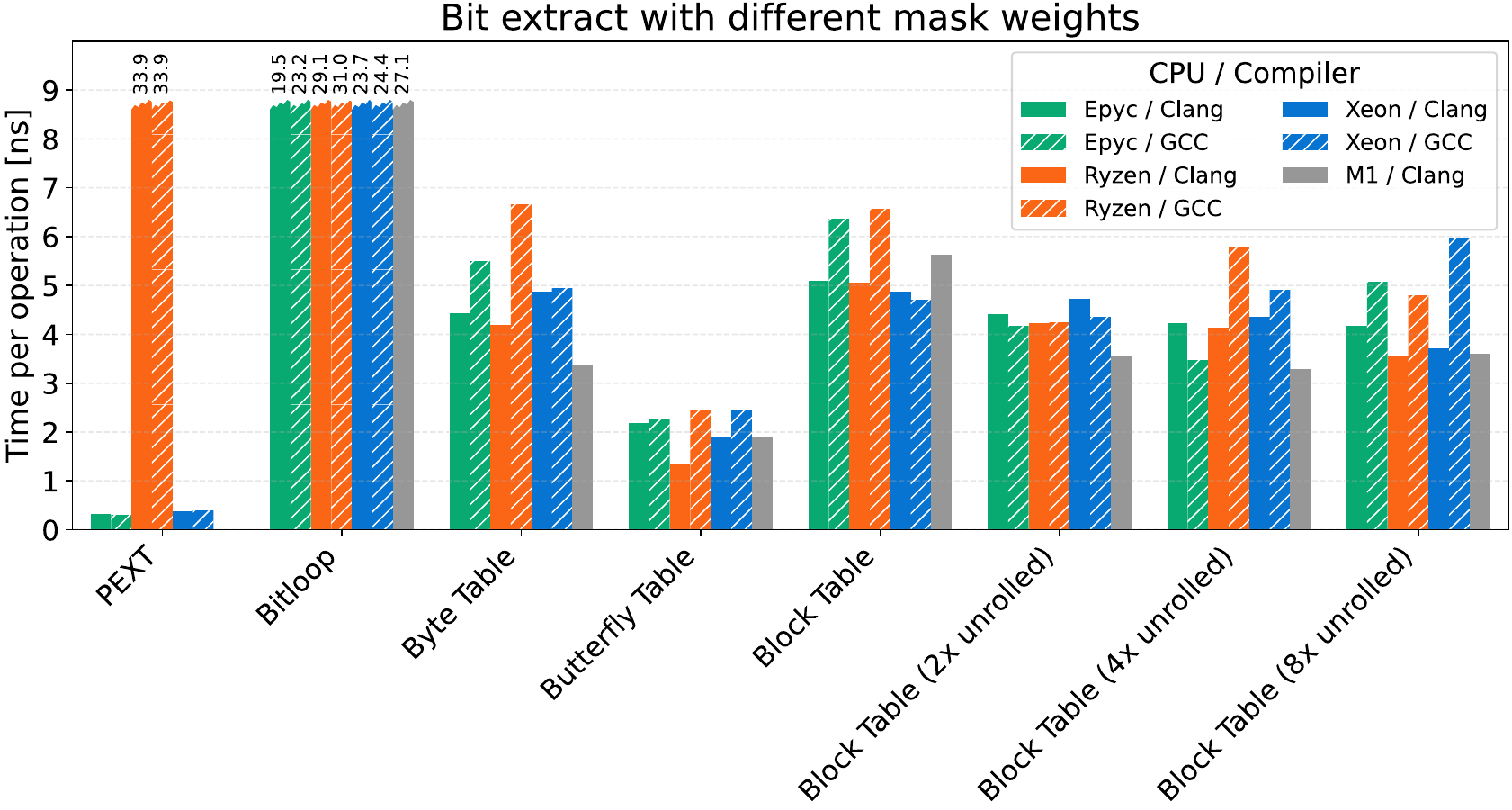}
    \\[1.5em]
    \includegraphics[width=0.8\linewidth]{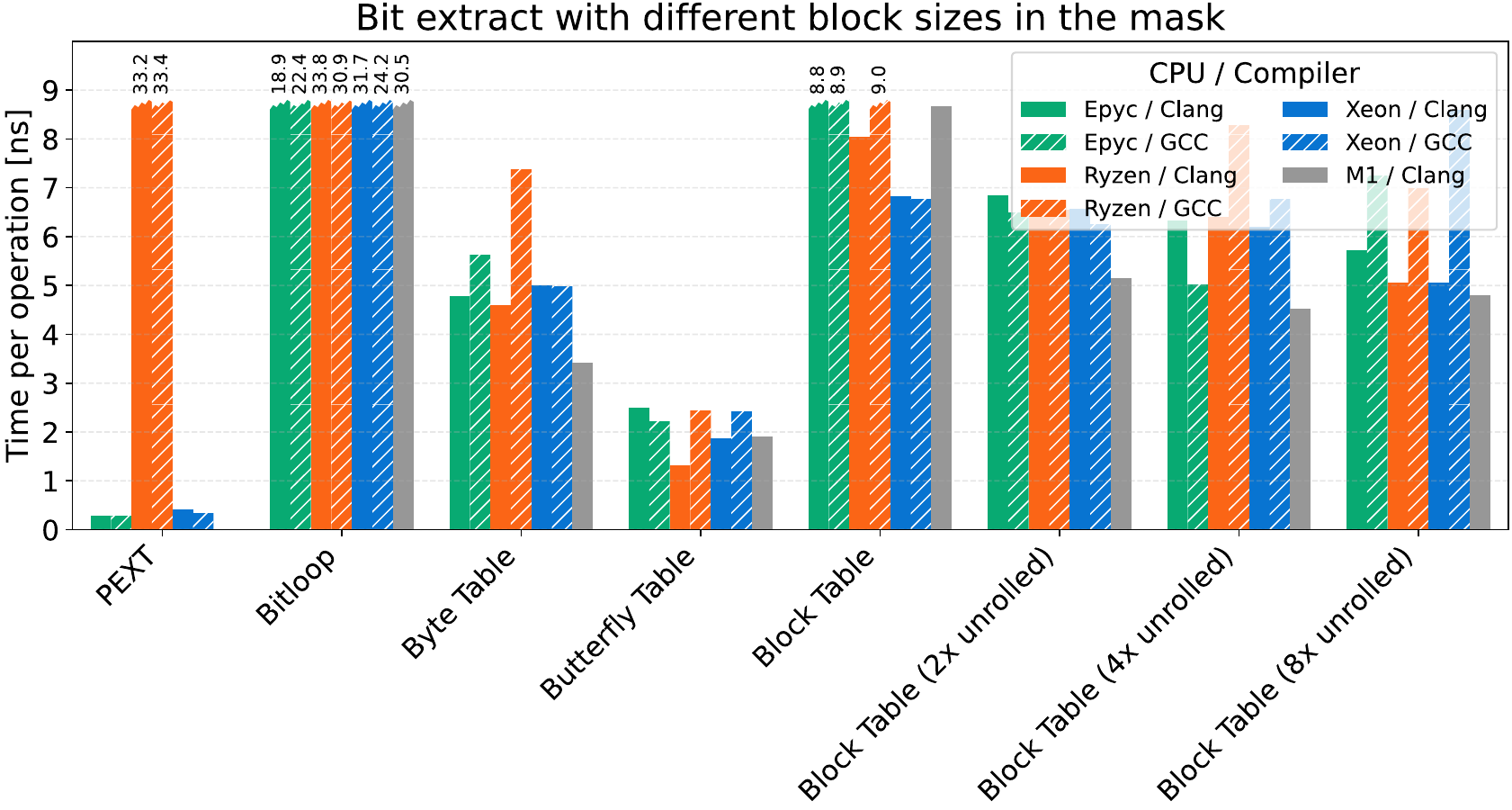}
    \end{tabular}
    \caption{
        \textbf{Bit extraction performance for different weights (top), and for different number of blocks (bottom), across architectures and compilers.}
        This shows the same underlying data as \figref{supp:fig:BitExtractWeights} and \figref{supp:fig:BitExtractBlocks}, but as a summary of all tested architectures and compilers. It shows the averages across all cases, i.\,e., all weights of the mask and all number of blocks, respectively. Here, the advantage of the \texttt{PEXT} hardware instruction is clearly visible, while the Butterfly network table is the most performant software implementation.
    }
\label{supp:fig:BitExtractSummary}
\end{figure}

\paragraph{Application.}
The application to an input value $x$ is shown in \algoref{alg:bit-extract-butterfly}. For each step, corresponding to a power of two, those bits are shifted whose decomposition contains that power.
As the number of stages is fixed for a given bit width, the loop can be trivially unrolled. However, as the stages depend on each other in their execution order, parallelism through instruction pipelining is somewhat limited. Still, this is generally the most efficient software implementation we tested, see \figref{supp:fig:BitExtractWeights}--\ref{supp:fig:BitExtractSummary} for the benchmarks.



\begin{algorithm}
\caption{Bit extraction using a precomputed butterfly table.}
\label{alg:bit-extract-butterfly}
\begin{algorithmic}[1]
\Function{bit-extract-butterfly}{$x, m, T$}
    \State $x \gets x ~\operatorname{AND}~ m$
        \Comment{Discard bits not selected by the original mask}

    \For{$i \gets 0$ \textbf{to} $5$} \Comment{Process all 6 rows of the table}
        \State $s \gets 2^i$ \Comment{Current shift distance (power of two)}

        \State $t \gets x ~\operatorname{AND}~ T[i]$
            \Comment{Extract bits that need to move at this stage}
        \State $x \gets (x ~\operatorname{XOR}~ t) ~\operatorname{OR}~ (t ~\text{\texttt{>>}}~ s)$
            \Comment{Remove them and insert them shifted right}
    \EndFor

    \State \Return $x$
\EndFunction
\end{algorithmic}
\end{algorithm}


\clearpage

\section{Spaced \textit{k}-mer Extraction}
\label{sec:SpacedExtraction}

We can now apply the generic bit extraction to the input $k$-mers to obtain $w$-mers. Note that the generic functions operate on individual bits; our 2-bit encoding requires pairs of bits to be used instead. This has however no effect on the performance of the algorithms.

\subsection{Input Character Validity}

One of the main goals of spaced $k$-mers is to be able to skip invalid characters in the input, if they fall into one of the spaces of the mask, without having to skip the whole $k$-mer. To achieve this, we need to adapt our above $k$-mer iteration of \algoref{alg:kmer-shift} to track valid bases while traversing the input.

This can be implemented efficiently with a second 64-bit word, shifted in sync with the $k$-mer, and setting two bits to indicate character validity at the respective positions in the $k$-mer. A simple bit comparison with the mask then reveals if all non-masked nucleotides are valid for any given $k$-mer. The overhead for this check did not significantly influence performance in our tests: Assuming that most input characters are valid (from $\mathrm{N}$), the branch predictor can easily avoid slowdown in most cases. Note that this is in principle similar to the desired property of spaced $k$-mers to be able to tolerate mismatches with a reference genome; here however, we are solely operating based on input character validity.

We show this adaptation in \algoref{alg:spaced-kmer-extract}, taking a sequence $S$, a value for $k$, a set of masks $\mathcal{M}$, and a function $f$ to apply to each extracted valid $w$-mer. For the single-mask case, the inner loop over masks can be omitted.

\begin{algorithm}
\caption{Rolling spaced \(k\)-mer extraction.}
\label{alg:spaced-kmer-extract}
\begin{algorithmic}[1]
\Function{for-each-spaced-kmer}{$S, k, \mathcal{M}, f$}
    \State $B \gets 2^{2k} - 1$
        \Comment{Mask retaining the lowest \(2k\) bits}

    \State $x \gets 0$
        \Comment{Current rolling $k$-mer}
    \State $v \gets 0$
        \Comment{Current rolling validity bits}

    \For{$i \gets 0$ \textbf{to} $|S| - 1$} \Comment{Iterate all $k$-mers in $S$}
        \State $c \gets \operatorname{enc}(S[i])$ \Comment{Encode a single character}

        \State $q \gets \begin{cases} 
            \mathtt{0b11}, & \text{if } c < 4,\\
            \mathtt{0b00}, & \text{otherwise}
        \end{cases}$ \Comment{Set the validity bits}

        \State $x \gets ((x ~\texttt{<<}~ 2) ~\operatorname{AND}~ B) ~\operatorname{OR}~ (c ~\operatorname{AND}~ \mathtt{0b11})$
            \Comment{Shift in the encoded character}
        \State $v \gets ((v ~\texttt{<<}~ 2) ~\operatorname{AND}~ B) ~\operatorname{OR}~ q$
            \Comment{Shift in the validity bits}

        \ForAll{$m \in \mathcal{M}$}
            \If{$(v ~\operatorname{AND}~ m) = m$} \Comment{If all current characters are valid...}
                \State $w \gets \operatorname{extract}(x, m)$ \Comment{... extract the $w$-mer using the mask...}
                \State $f(w)$ \Comment{... and process it}
            \EndIf
        \EndFor
    \EndFor
\EndFunction
\end{algorithmic}
\end{algorithm}




\subsection{Single Mask}
\label{sec:SpacedExtraction:sub:SingleMask}

\begin{figure}[!htb]
    \centering
    \begin{tabular}{@{}cc@{}}
    \includegraphics[width=0.49\linewidth]{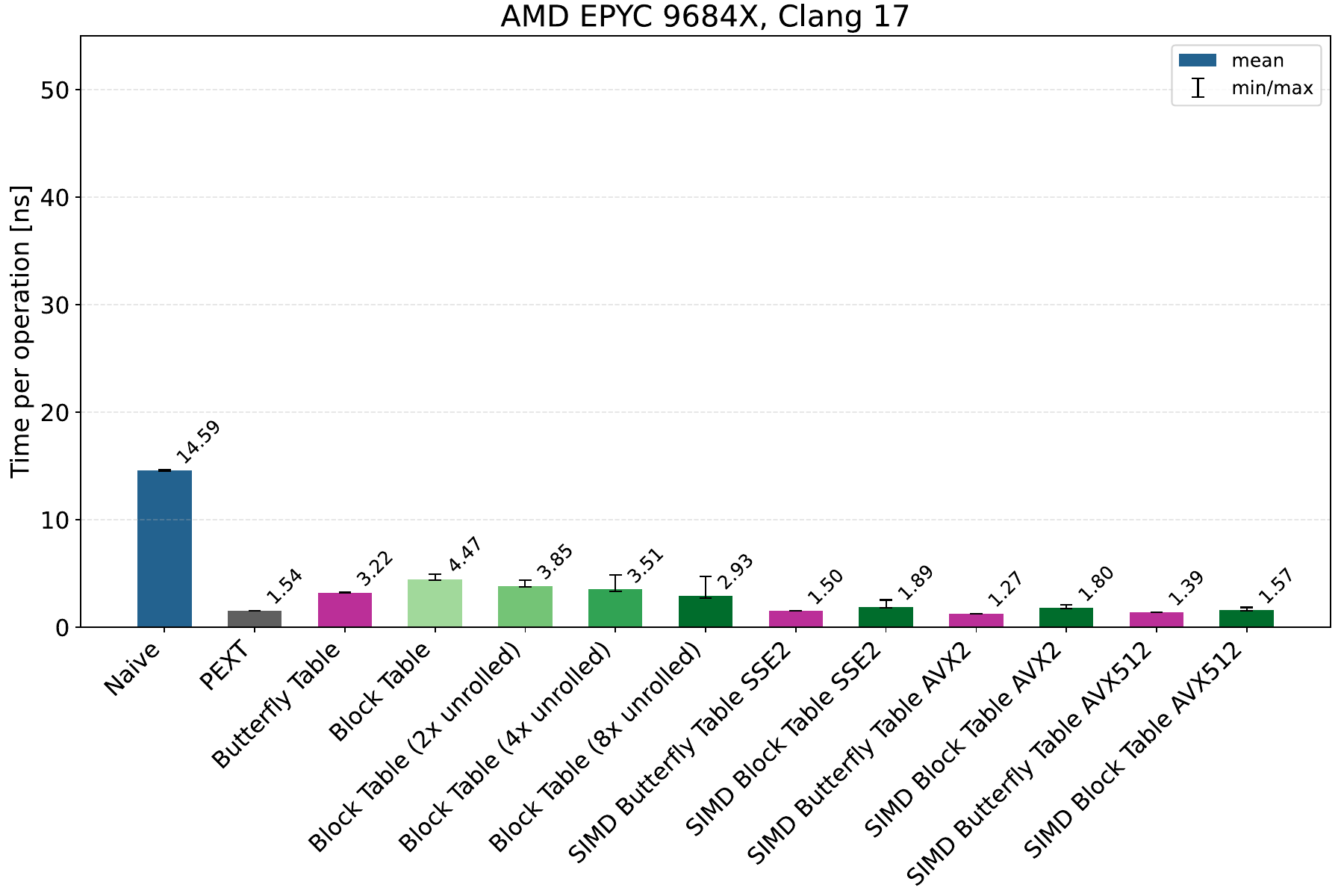} &
    \includegraphics[width=0.49\linewidth]{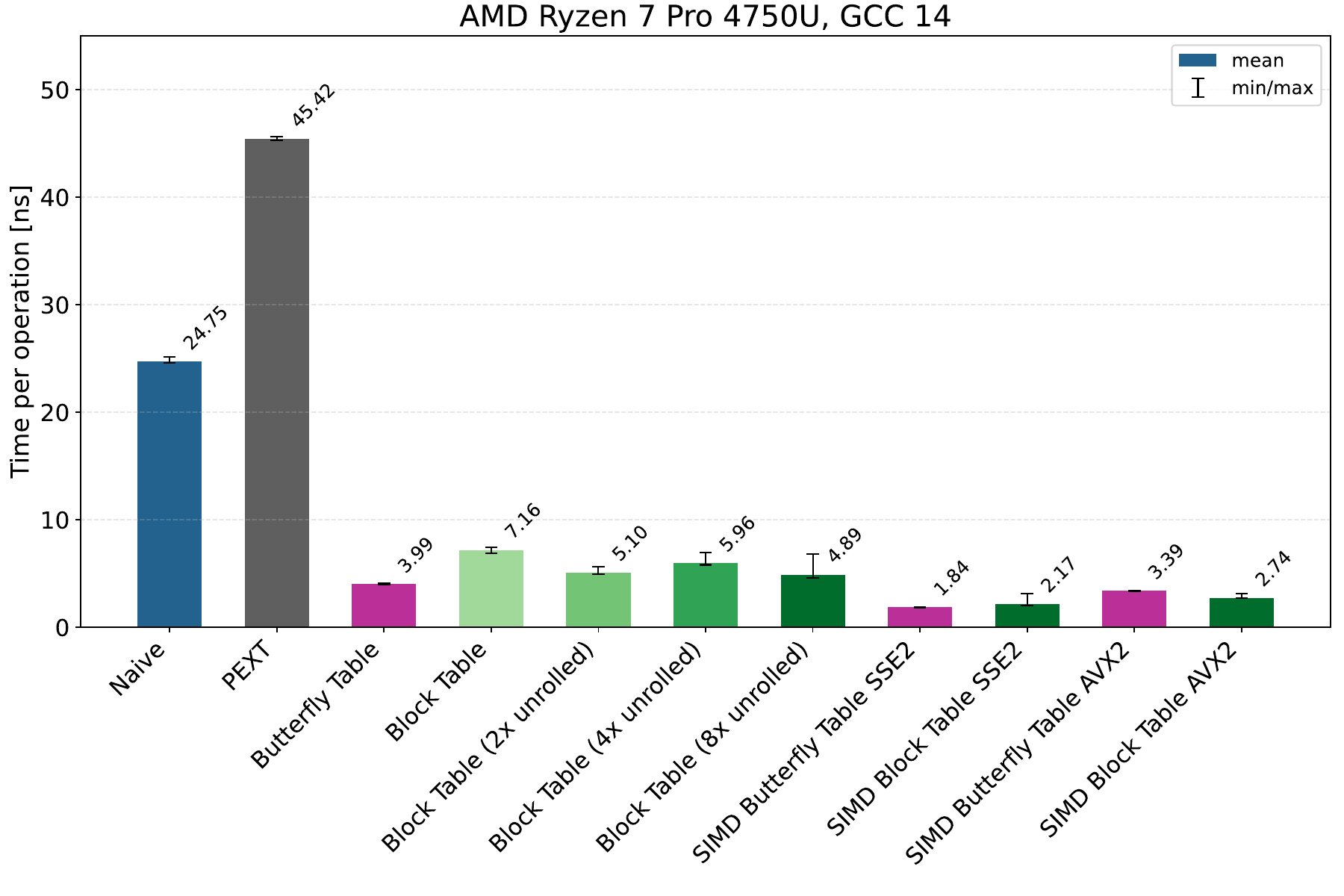}
    \\[0.5em]
    \includegraphics[width=0.49\linewidth]{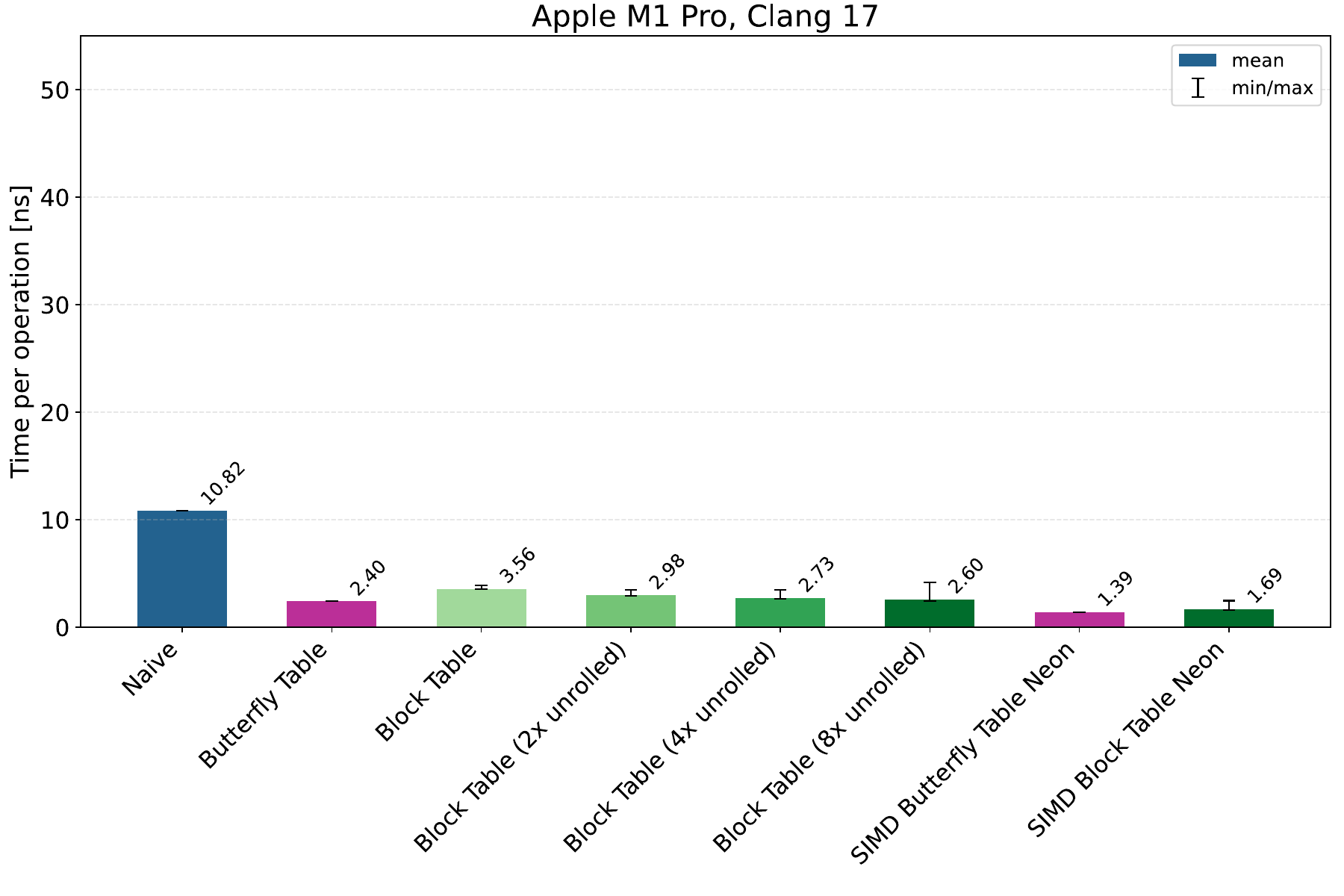} &
    \includegraphics[width=0.49\linewidth]{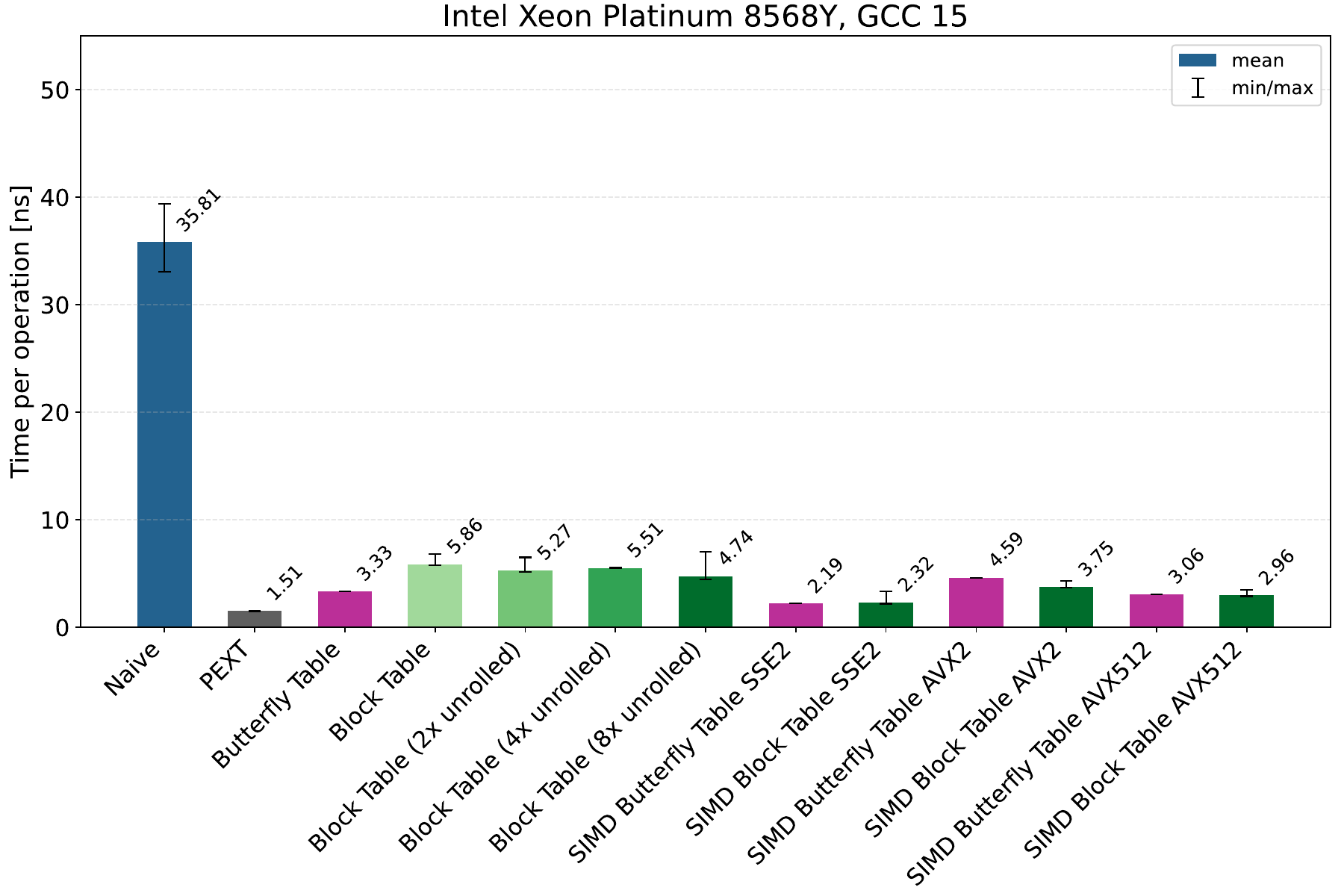}
    \end{tabular}
    \begin{subfigure}{0pt}
        \phantomcaption
        \label{fig:KmerSpacedSingleBars:sub:AMDEpyc9684XClang17}
    \end{subfigure}
    \begin{subfigure}{0pt}
        \phantomcaption
        \label{fig:KmerSpacedSingleBars:sub:AMDRyzen7Pro4750UGCC14}
    \end{subfigure}
    \begin{subfigure}{0pt}
        \phantomcaption
        \label{fig:KmerSpacedSingleBars:sub:AppleM1ProClang17}
    \end{subfigure}
    \begin{subfigure}{0pt}
        \phantomcaption
        \label{fig:KmerSpacedSingleBars:sub:IntelXeonPlatinum8568YGCC15}
    \end{subfigure}
    \vspace*{-1em}
    \caption{
        \textbf{Extraction speed of spaced $k$-mers for single masks.} This shows the time needed per $w$-mer to traverse an input sequence, including the $k$-mer processing and bit extraction to obtain the $w$-mer. Bars are the average across 9 distinct masks with weight $w=22$ and span $k=31$, and whiskers show the minimum and maximum across these masks.
    }
\label{supp:fig:KmerSpacedSingleBars}
\end{figure}

Benchmarks are shown in \figref{supp:fig:KmerSpacedSingleBars}; these are detailed variants of Figure 2(a) from the main manuscript, for a representative selection of hardware architectures. In all cases, there is at least one implementation that achieves processing one $w$-mer per 1.3--1.8\,ns. Note that the plots include SIMD accelerations as explained below.

\subsection{Multiple Masks}
\label{sec:SpacedExtraction:sub:MultipleMasks}

Many applications of spaced $k$-mers use a set of distinct masks in order to increase sensitivity. With multiple masks, the bit extraction is simply applied to each $k$-mer and for each mask individually, leading to a general increase in runtime per additional mask. However, the nucleotide encoding and $k$-mer iteration only has to be performed once then, which amortizes the fraction of runtime for these tasks, and leading to an increase in performance when measuring throughput per distinct mask and $w$-mer. For ease of comparison with the single mask case, we report this time per $w$-mer here; note though that the runtime per $k$-mer thus needs to be multiplied by the number of masks used (here, 9 masks).

\begin{figure}[!htb]
    \centering
    \begin{tabular}{@{}cc@{}}
    \includegraphics[width=0.49\linewidth]{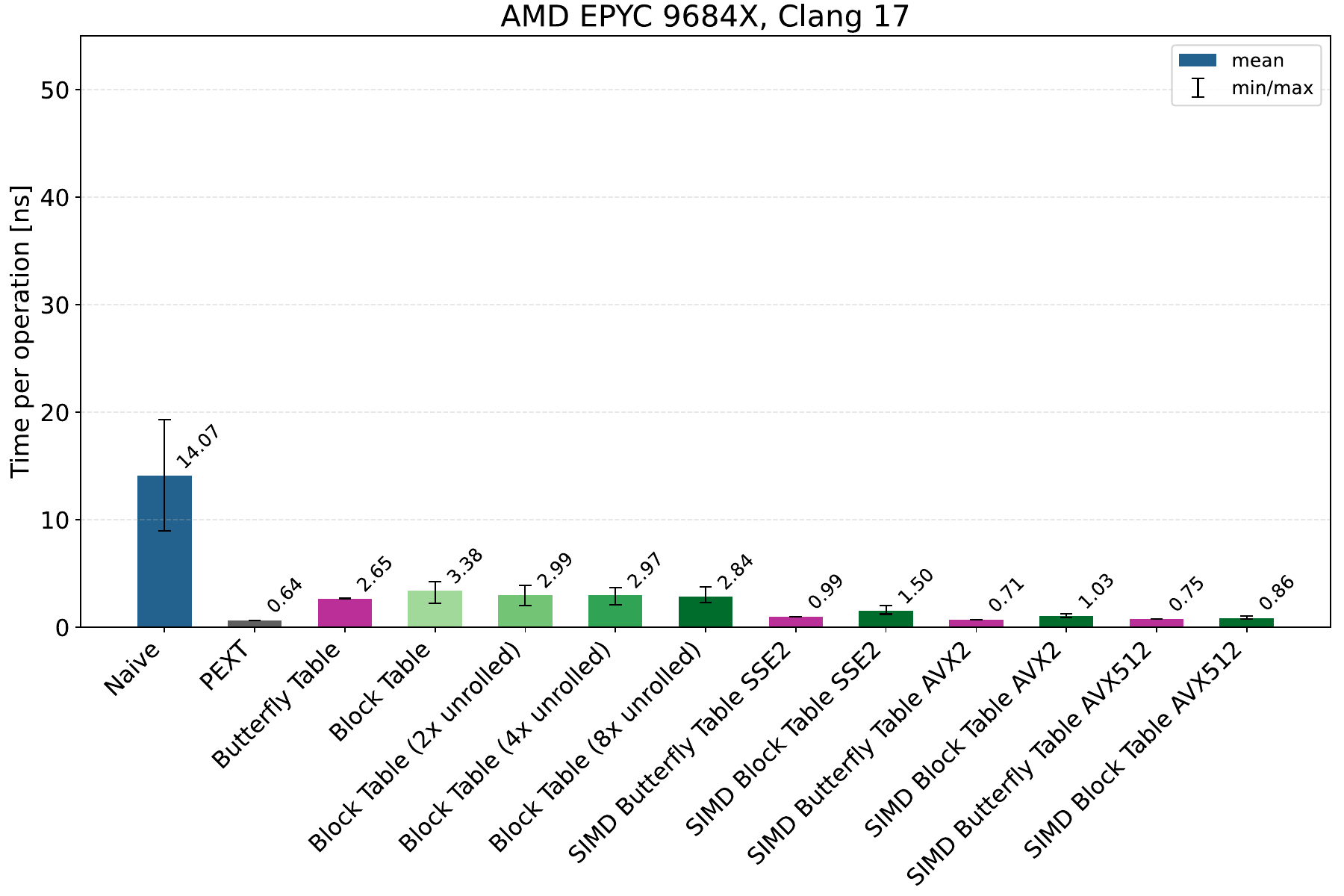} &
    \includegraphics[width=0.49\linewidth]{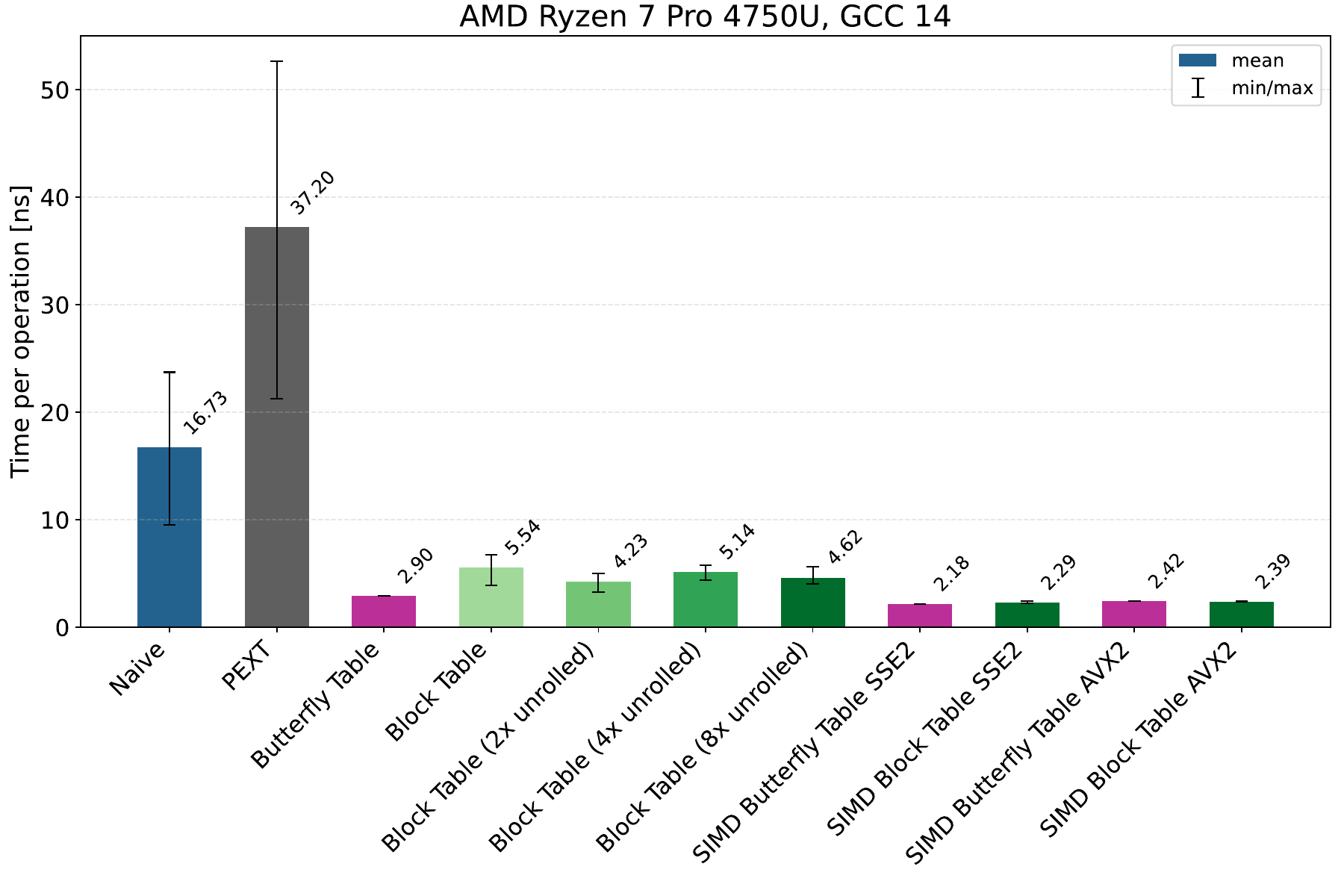}
    \\[0.5em]
    \includegraphics[width=0.49\linewidth]{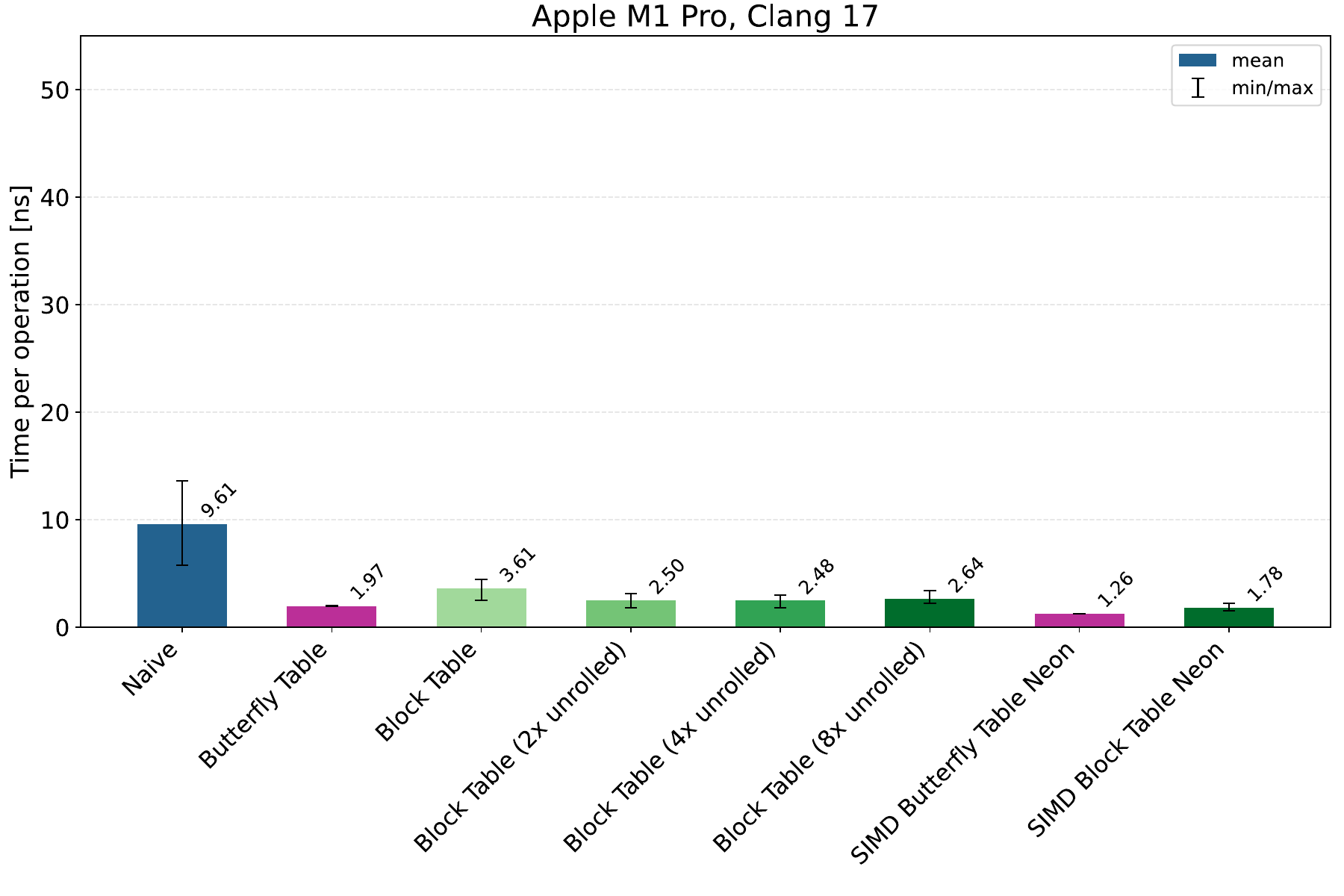} &
    \includegraphics[width=0.49\linewidth]{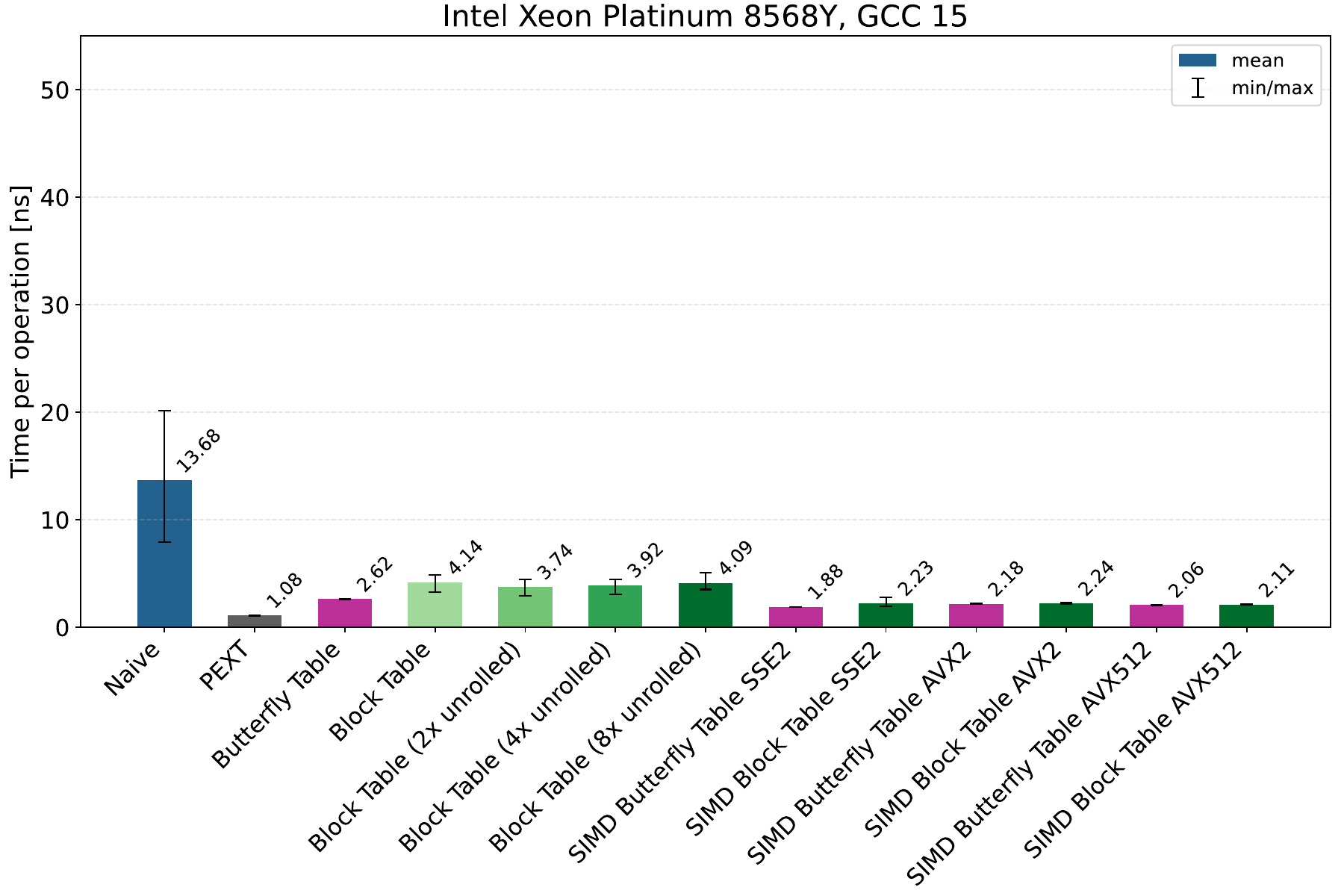}
    \end{tabular}
    \begin{subfigure}{0pt}
        \phantomcaption
        \label{fig:KmerSpacedMultiBars:sub:AMDEpyc9684XClang17}
    \end{subfigure}
    \begin{subfigure}{0pt}
        \phantomcaption
        \label{fig:KmerSpacedMultiBars:sub:AMDRyzen7Pro4750UGCC14}
    \end{subfigure}
    \begin{subfigure}{0pt}
        \phantomcaption
        \label{fig:KmerSpacedMultiBars:sub:AppleM1ProClang17}
    \end{subfigure}
    \begin{subfigure}{0pt}
        \phantomcaption
        \label{fig:KmerSpacedMultiBars:sub:IntelXeonPlatinum8568YGCC15}
    \end{subfigure}
    \caption{
        \textbf{Extraction speed of spaced $k$-mers for multiple masks.} This shows the average time needed per $w$-mer to traverse an input sequence, including the $k$-mer processing and bit extraction to obtain all $w$-mers for the given set of masks. Bars are the average across 5 distinct mask sets with 9 masks each, with span and weight $(k,w) \in \{ (15,10), (31,14), (31, 18), (31,22), (31,26) \}$ for the sets, and whiskers show the minimum and maximum across the mask sets. Note that for comparability with the single mask case above, this shows the average time per $w$-mer; for the complete time to extract all 9 $w$-mers for a given mask set, values have hence to be multiplied by 9.
    }
\label{supp:fig:KmerSpacedMultiBars}
\end{figure}

Benchmarks for 5 distinct sets of 9 masks each are shown in \figref{supp:fig:KmerSpacedMultiBars}. In the best case, on Intel Xeon, we achieve 0.58\,ns per $w$-mer using the PEXT instruction.
The plots again include SIMD accelerations as explained next.


\subsection{SIMD Acceleration}
\label{sec:SpacedExtraction:sub:SIMD}

The generic algorithms for the \texttt{bit extract} operation as introduced above were so far applied independently on each $k$-mer to extract the $w$-mer. However, in our typical use case, we are traversing a sequence, applying an extraction with the same mask to consecutive $k$-mers. To accelerate this, we can exploit SIMD (single instruction, multiple data) intrinsics.

The two most performant implementations, block table and butterfly network, are particularly suitable for SIMD parallelization: They only need a small set of pre-computed values per mask, and can efficiently apply them in parallel across multiple 64-bit lanes. We have implemented both approaches for several SIMD instruction sets. These operate on different underlying register sizes, and hence differ in the number of lanes they can process at once. In particular, we implemented both algorithms for SSE2 (128-bit; 2 lanes), AVX2 (256-bit; 4 lanes), AVX512 (512-bit, 8 lanes), and ARM Neon (128-bit, 2 lanes). The main loop over the input sequence then extracts as many consecutive $k$-mers as available lanes, and extracts $w$-mers in the lanes in parallel.

\begin{figure}[!htb]
    \centering
    \includegraphics[width=\linewidth]{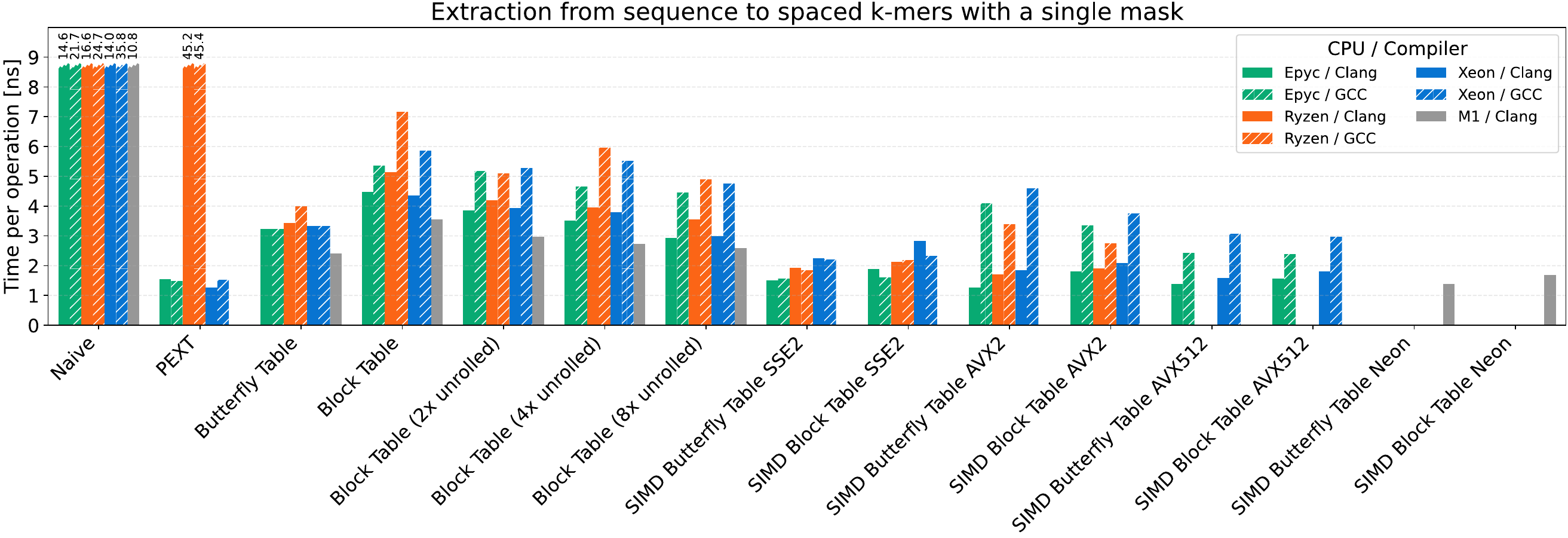}
    \includegraphics[width=\linewidth]{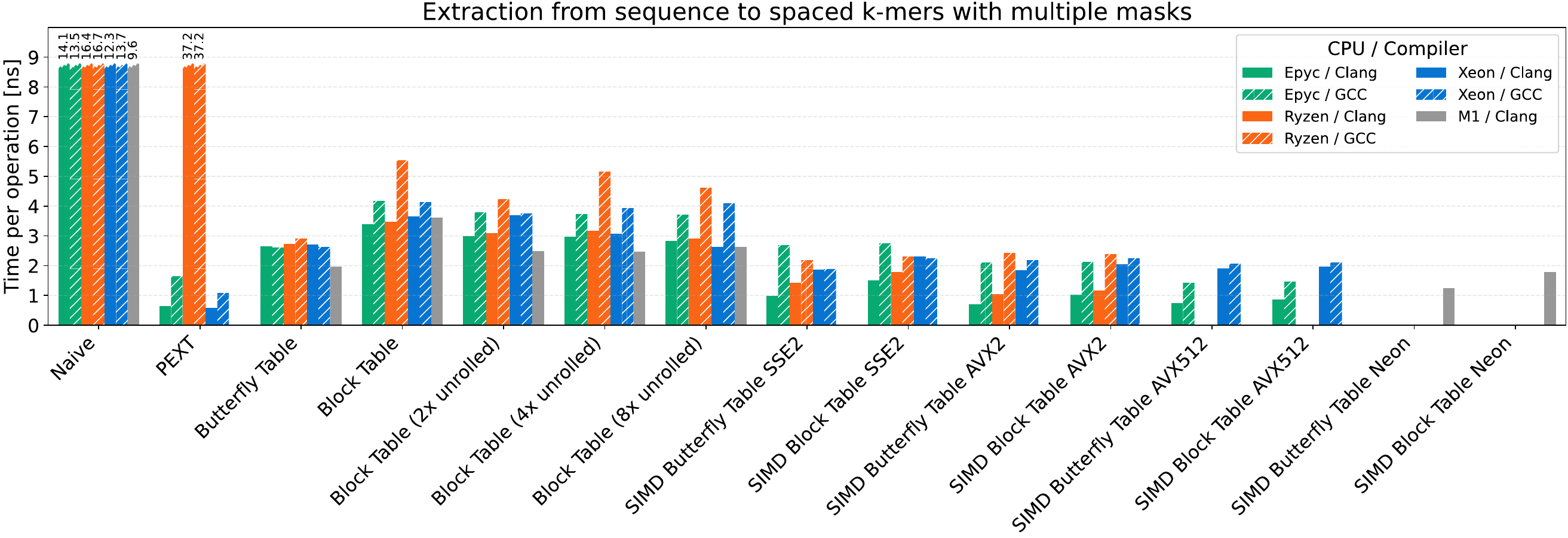}
    \caption{
        \textbf{Extraction speed of spaced $k$-mers for single (top) and multiple (bottom) masks.} This shows the same underlying data as \figref{supp:fig:KmerSpacedSingleBars} and \figref{supp:fig:KmerSpacedMultiBars}, but summarized across all architectures and compilers. It is an extension of Figure 2 in the main manuscript, containing benchmarks of all implementations presented here.
    }
\label{supp:fig:KmerSpacedSummary}
\end{figure}

Exemplary benchmarks are included in the previously described \figref{supp:fig:KmerSpacedSingleBars} and \figref{supp:fig:KmerSpacedMultiBars}; a summary across all architectures and compilers is provided in \figref{supp:fig:KmerSpacedSummary} and in \tabref{tab:SummarySpeeds}.

\begin{table}[]
\centering
\tabcolsep=0.15cm
\begin{tabular}{l|lrr|lrrr}
               & \multicolumn{3}{c|}{Single mask}                                          & \multicolumn{4}{c}{Multiple masks}                                                                   \\
CPU / Compiler & Fastest algorithm & \multicolumn{1}{l}{Time} & \multicolumn{1}{l|}{Speed} & Fastest algorithm & \multicolumn{1}{l}{Time} & \multicolumn{1}{l}{Total} & \multicolumn{1}{l}{Speed} \\ \hline
Epyc / Clang   & Butterfly AVX2    & 1.27                     & 750                        & PEXT              & 0.64                     & 5.72                      & 167                       \\
Epyc / GCC     & PEXT              & 1.47                     & 648                        & Butterfly AVX512  & 1.41                     & 12.68                     & 75                        \\
Ryzen / Clang  & Butterfly AVX2    & 1.71                     & 557                        & Butterfly AVX2    & 1.04                     & 9.38                      & 102                       \\
Ryzen / GCC    & Butterfly SSE2    & 1.84                     & 517                        & Butterfly SSE2    & 2.18                     & 19.59                     & 49                        \\
Xeon / Clang   & PEXT              & 1.28                     & 747                        & PEXT              & 0.58                     & 5.21                      & 183                       \\
Xeon / GCC     & PEXT              & 1.51                     & 631                        & PEXT              & 1.08                     & 9.71                      & 98                        \\
M1 / Clang     & Butterfly Neon    & 1.39                     & 687                        & Butterfly Neon    & 1.26                     & 11.30                     & 84                       
\end{tabular}
\caption{\textbf{Performance of the fastest algorithm across hardware architectures.} The table summarizes the above benchmarks, showing the time (ns per $w$-mer) and speed (MB/s) for the single mask case, as well as for the multiple mask case the total time per input character (ns per $k$-mer) for all 9 masks combined and the resulting speed. Note that typical applications usually use fewer masks; the speed here is thus exemplary only.}
\label{tab:SummarySpeeds}
\end{table}


\subsection{Dynamic Algorithm Selection}
\label{sec:BitExtract:sub:DynamicAlgorithmSelection}

\paragraph{Hardware availability.}
Except for the basic implementations, our accelerations depend on hardware availability of CPU intrinsics: \texttt{PEXT} for the fastest bit extraction, as well as SSE2, AVX2, AVX512, and ARM Neon for our SIMD implementations of the Block table and Butterfly algorithms. While these give significant speed improvements, not all instruction sets are available on different hardware architectures. Hence, typically, code needs to be compiled per architecture. 
Unfortunately, however, the \texttt{PEXT} instruction is available on older AMD CPUs, but very inefficient, as shown above. Hence, simply testing for the underlying BMI2 instruction set at compile time could lead to significant performance degradation.

\paragraph{Runtime dispatch.}
Thus, we use a runtime dispatch between the different algorithms, and switch to the fastest available implementation. To facilitate this, we offer a mechanism for dynamically selecting algorithms, which runs a short benchmark on random input. This is executed for a fixed mask, which is sufficient for most typical use cases for spaced $k$-mers. The mechanism can be readily implemented in applications that wish to use our code base. For completeness, we offer a similar dynamic selection for the generic bit extract functionality. 

\paragraph{Cross compilation.}
These mechanisms have the additional advantage of allowing cross compilation. As the algorithm is selected once dynamically at the beginning of the program runtime, an additional check can be performed to test if any given intrinsic is actually available on the current CPU, thus avoiding to crash the program with invalid instructions. We implemented this in our selection mechanism. This is particularly useful to compile a program for platforms such as conda, where a single binary can then serve any hardware architecture. 

\paragraph{Inlining.}
To minimize overhead for dispatch in the hot loop, and to allow function inlining by the compiler, we recommend to use the dynamic selection to dispatch once in an outer context, using C++ templates for the inner function. We provide exemplary code for this.

\subsection{Differences Between Compilers}
\label{sec:SpacedExtraction:sub:Compilers}

As can be seen in the benchmarks above, the performance in some cases has significant differences between compilers, even on the same hardware architecture. This is not surprising, as Clang and GCC tend to inline and optimize code differently. It is however out of scope of this manuscript to explore this in more detail. In any practical application of the presented methods, the spaced $k$-mer extration is typically followed by some downstream work anyway, which will cause compilers to optimize loops and functions in unpredictable ways. We thus recommend to test performance of applications under different compilers.


\subsection{Comparison With Existing Methods}
\label{supp:sec:MethodDetails:sub:ExistingMethods}


We compare our approach to the existing methods FSH \cite{Girotto2018-ns}, FISH \cite{Girotto2018-wt}, ISSH \cite{Petrucci2020-zt}, MISSH \cite{Mian2024-nf}, and DuoHash \cite{Gemin2026-du}. We here call these approaches the Comin \& Pizzi family of methods.

\paragraph{FSH, ISSH, and MISSH.}
These methods are build on the idea of re-using information across $w$-mers of previous positions along the input sequence. Using the auto-correlation of the bits in the mask with its shifted self, the methods find overlapping unmasked blocks of nucleotides of the underlying sequence. These are then re-used to fill the current $w$-mer based on previously observed ones, thus avoiding to re-encode characters from the underlying sequence. The methods are based on the assumption that encoding the characters into their 2-bit representation is expensive. In their original implementation, this was indeed the case, as they used an expensive series of \texttt{if}-statements, as explained in Section \ref{sec:EncodingIteration:sub:Encoding}. Using this approach as their baseline for comparison allowed them to report large performance improvements. They have since updated this, with consequences as explained below.

\paragraph{FISH and DuoHash.}
These methods use blocks of consecutive 1s in the mask to accelerate spaced $k$-mer extraction via  lookup tables. FISH has some conceptual similarity with our block table algorithm, but instead of simply shifting blocks, it uses a set of lookup tables for different block sizes. DuoHash similarly uses large lookup tables to re-use previously observed blocks of the input sequence. This makes DuoHash more general than simple encoding of the $k$-mers, as it can be used for computing actual hash values of spaced $k$-mers, which they showcase via the ntHash method \cite{Mohamadi2016-gg,Kazemi2022-nk}. We however did not test the DuoHash method here, as the code repository at \href{https://github.com/CominLab/DuoHash}{github.com/CominLab/DuoHash} does not seem to contain an actual implementation of the method (at the time of writing), and instead contains a concise interface for all other methods of the family. It is however an interesting approach that is useful in applications where spaced $k$-mers are further processed into hash values for lookup tables.

\paragraph{Overhead.}
These methods require a large amount of bookkeeping to manage block overlaps and lookup tables, which introduces functional complexity, as well as pointer indirections to access their data structures. Furthermore, the overlap-based methods require to keep a list of previous $w$-mers during iteration, causing some slowdown due to allocations and memory accesses. In fact, DuoHash \cite{Gemin2026-du} states that ``the speed-up generally tends to decrease slightly as the length of the reads increases'', which can be explained by this. In their current implementation, they store the previous $w$-mers for the whole sequence, despite a short window sufficing; this could hence be optimized.

\paragraph{Benchmark program.}
To compare our presented algorithms with the Comin \& Pizzi family of methods, we implemented a benchmark for their methods in our fork of their DuoHash repository at \href{https://github.com/lczech/DuoHash}{github.com/ lczech/DuoHash}. There, we added a simple program to the original code that runs comparable benchmarks to the ones we showed above, measuring the time per $k$-mer for a randomly generated input string. 

\begin{figure}[!bh]
    \centering
    \includegraphics[scale=0.29,valign=t]{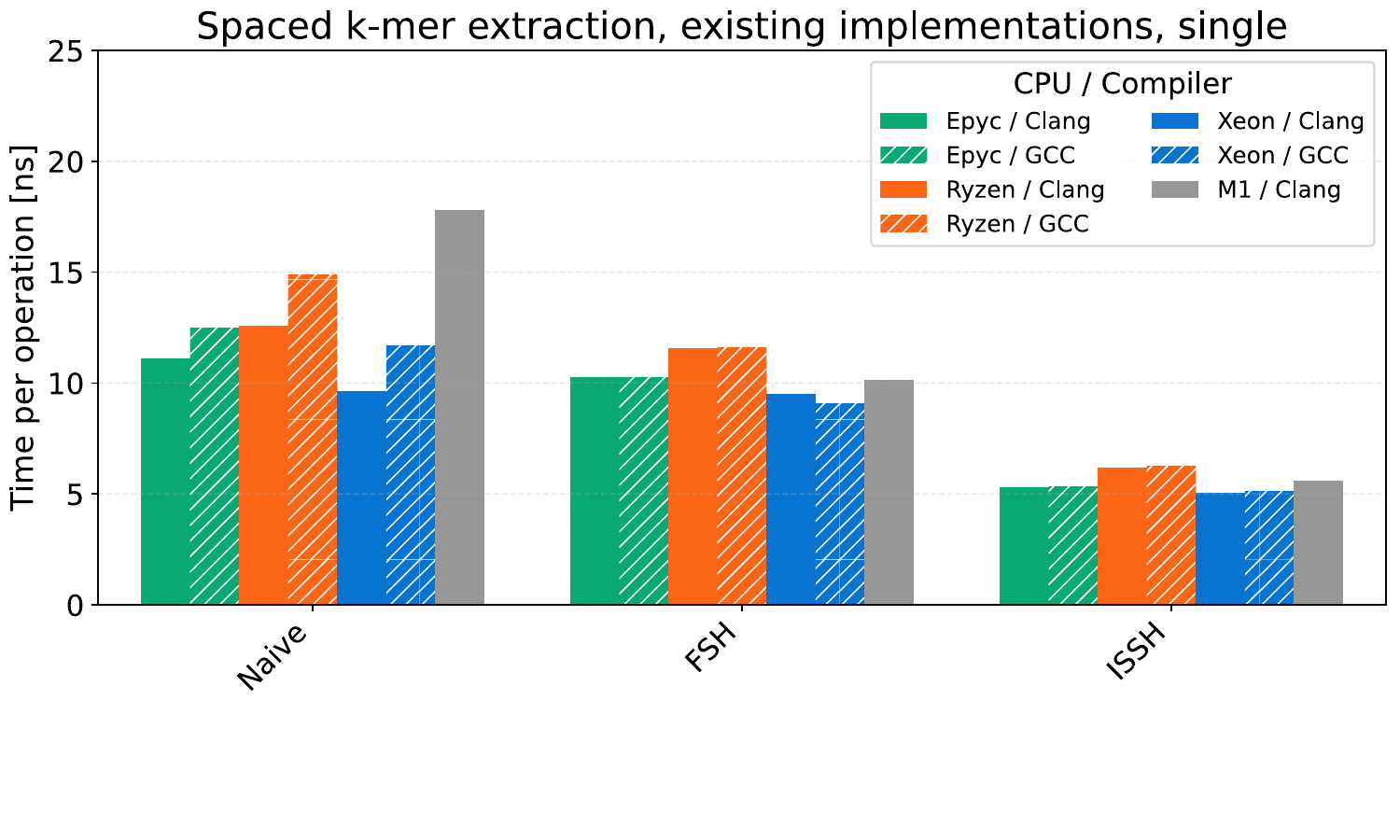}
    \includegraphics[scale=0.29,valign=t]{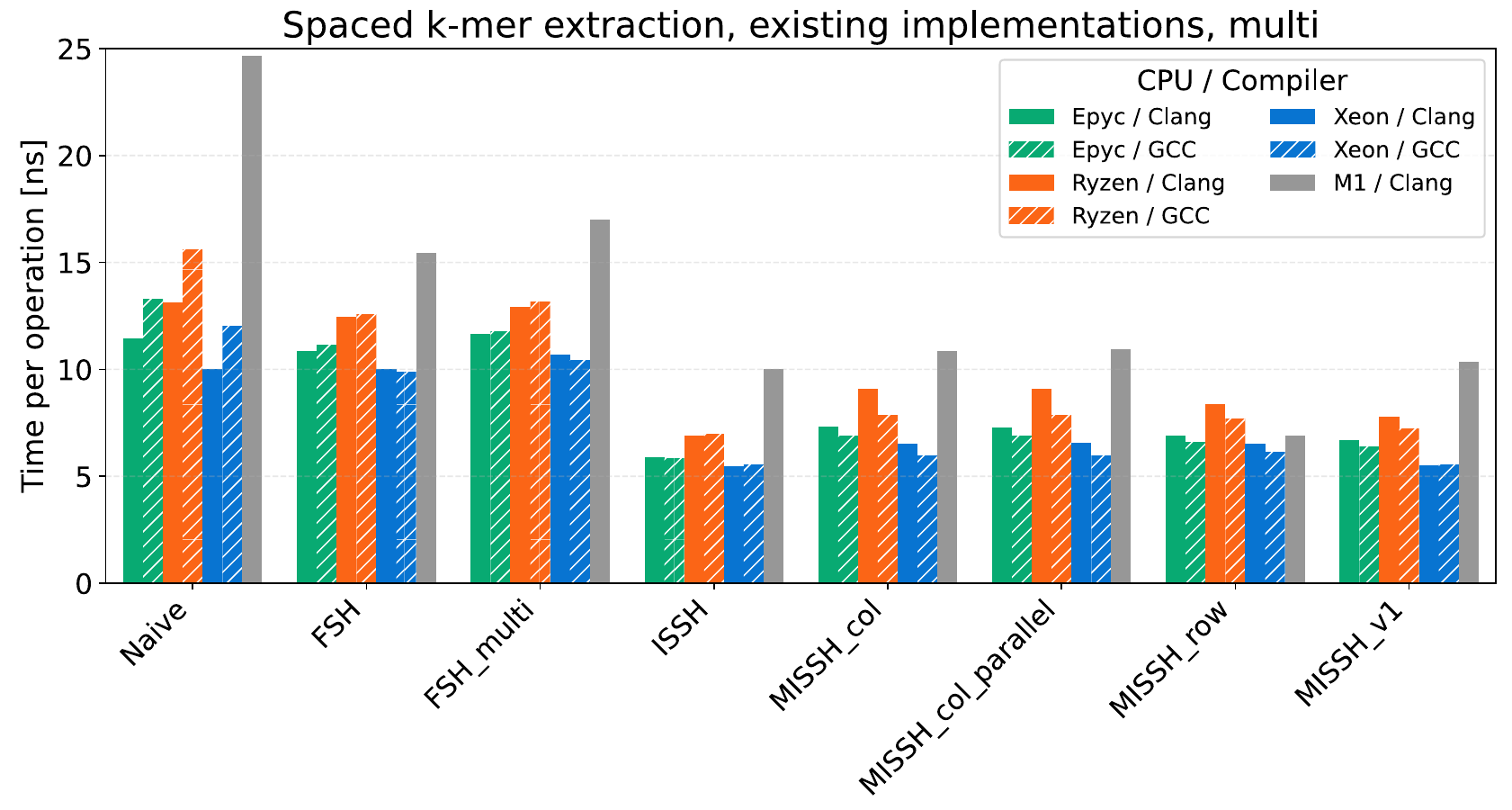}
    \caption{
        \textbf{Processing speed of spaced $k$-mers with existing methods of the Comin \& Pizzi family.} We implemented a benchmark program using the suite of existing approaches and their existing implementations, similar to our benchmarks above. The figure shows average time for processing a spaced $k$-mer on randomly generated data, as before. Across most architectures, the fastest method is ISSH, which is however still significantly slower than our fastest algorithm on the same architecture, respectively, and only about twice as fast as their  naive baseline implementation.
    }
\label{supp:fig:DuoHash}
\end{figure}

\paragraph{Implementation changes.}
The DuoHash code repository has re-implemented their benchmarking harness compared to the original one used in MISSH at \href{https://github.com/CominLab/MISSH}{github.com/CominLab/MISSH}. This update has one particular change that significantly improved the performance of their methods: Previously, as mentioned, they used a series of \texttt{if}-statements to encode the input characters into 2-bit encoding, while in their update this is implemented as a fast lookup table. This thus eliminates the bottleneck due to the encoding, diminishing the gains achieved with these methods compared to the naive baseline implementation, as shown in \figref{supp:fig:DuoHash}.

\paragraph{Comparison.}
Across almost all tested architectures, ISSH is the fastest of the Comin \& Pizzi methods, except on the Apple M1, where MISSH\_row is relatively the fastest (but absolutely, slower than on the x86 architecture). In the best case, the methods reached a throughput of 5.0\,ns per extracted $w$-mer.
In comparison, out of our algorithms presented here, PEXT is the fastest where available in hardware, and one of the SIMD Butterfly implementations otherwise. Across hardware architectures, our algorithms for a single mask are 3.4--4.2 times faster, and for multiple masks 3.2--9.5 times faster, than the best out of the Comin \& Pizzi methods on the respective architecture. Additionally, our basic algorithms are also significantly easier to implement (the SIMD acceleration adds some complexity though), and do not need to keep the list of previous $k$-mers in memory during the iteration.

\paragraph{Limitations and future work.}
Note that these comparisons are limited to $k \leq 32$. Adapting our algorithms to larger values of $k$ would require adaptation to 128-bit registers, or introducing additional bit operations to work across 64-bit boundaries. The former approach would allow $k \leq 64$, while the latter has the advantage that it could work with arbitrary sizes of $k$. It is left as future work to explore such algorithms, and the implications for performance. The Comin \& Pizzi methods can work with $k > 32$, but are still currently limited to $w \leq 32$, as they also store the $w$-mers in 64-bit words. Adapting them to larger values of $w$ would likely introduce similar additional bit operations as would be needed for our algorithms. We hence consider it is likely that adaptations of our algorithms to $k > 32$ would still outperform the Comin \& Pizzi ones.






\pagebreak
\clearpage

\bibliographystyle{pnas-new}

\bibliography{references_supp}